\documentclass[twocolumn,aps,prd,floatfix,nofootinbib,superscriptaddress]{revtex4-2}
\usepackage[utf8]{inputenc}
\usepackage{amssymb}
\usepackage{graphicx}
\usepackage{import}
\usepackage{mathtools}
\usepackage{hyperref}
\usepackage{mathrsfs}
\usepackage{multirow}
\usepackage{epstopdf}
\usepackage{longtable}

\def\mbb{\linebreak[0]}

\begin{document}

\title{\boldmath Study of the process \( e^+e^- \to \omega\pi^0 \to \pi^+\pi^-\pi^0\pi^0 \) in the energy range 1.05--2.00 GeV with SND}

\author{M.~N.~Achasov}
\affiliation{Budker Institute of Nuclear Physics, SB RAS, Novosibirsk, 630090, Russia}
\affiliation{Novosibirsk State University, Novosibirsk, 630090, Russia}
\author{A.~Yu.~Barnyakov}
\affiliation{Budker Institute of Nuclear Physics, SB RAS, Novosibirsk, 630090, Russia}
\affiliation{Novosibirsk State University, Novosibirsk, 630090, Russia}
\affiliation{Novosibirsk State Technical University, Novosibirsk, 630073, Russia}
\author{K.~I.~Beloborodov}
\affiliation{Budker Institute of Nuclear Physics, SB RAS, Novosibirsk, 630090, Russia}
\affiliation{Novosibirsk State University, Novosibirsk, 630090, Russia}
\author{A.~V.~Berdyugin}
\affiliation{Budker Institute of Nuclear Physics, SB RAS, Novosibirsk, 630090, Russia}
\affiliation{Novosibirsk State University, Novosibirsk, 630090, Russia}
\author{D.~E.~Berkaev}
\affiliation{Budker Institute of Nuclear Physics, SB RAS, Novosibirsk, 630090, Russia}
\affiliation{Novosibirsk State University, Novosibirsk, 630090, Russia}
\author{A.~G.~Bogdanchikov}
\affiliation{Budker Institute of Nuclear Physics, SB RAS, Novosibirsk, 630090, Russia}
\author{A.~A.~Botov}
\affiliation{Budker Institute of Nuclear Physics, SB RAS, Novosibirsk, 630090, Russia}
\author{V.~S.~Denisov}
\affiliation{Budker Institute of Nuclear Physics, SB RAS, Novosibirsk, 630090, Russia}
\author{T.~V.~Dimova}
\affiliation{Budker Institute of Nuclear Physics, SB RAS, Novosibirsk, 630090, Russia}
\affiliation{Novosibirsk State University, Novosibirsk, 630090, Russia}
\author{V.~P.~Druzhinin}
\affiliation{Budker Institute of Nuclear Physics, SB RAS, Novosibirsk, 630090, Russia}
\affiliation{Novosibirsk State University, Novosibirsk, 630090, Russia}
\author{E.~A.~Eminov}
\affiliation{Budker Institute of Nuclear Physics, SB RAS, Novosibirsk, 630090, Russia}
\author{L.~B.~Fomin}
\affiliation{Budker Institute of Nuclear Physics, SB RAS, Novosibirsk, 630090, Russia}
\author{L.~V.~Kardapoltsev}
\affiliation{Budker Institute of Nuclear Physics, SB RAS, Novosibirsk, 630090, Russia}
\affiliation{Novosibirsk State University, Novosibirsk, 630090, Russia}
\author{A.~G.~Kharlamov}
\affiliation{Budker Institute of Nuclear Physics, SB RAS, Novosibirsk, 630090, Russia}
\affiliation{Novosibirsk State University, Novosibirsk, 630090, Russia}
\author{I.~A.~Koop}
\affiliation{Budker Institute of Nuclear Physics, SB RAS, Novosibirsk, 630090, Russia}
\affiliation{Novosibirsk State University, Novosibirsk, 630090, Russia}
\author{A.~A.~Korol}
\affiliation{Budker Institute of Nuclear Physics, SB RAS, Novosibirsk, 630090, Russia}
\affiliation{Novosibirsk State University, Novosibirsk, 630090, Russia}
\author{D.~P.~Kovrizhin}
\affiliation{Budker Institute of Nuclear Physics, SB RAS, Novosibirsk, 630090, Russia}
\author{A.~S.~Kupich}
\affiliation{Budker Institute of Nuclear Physics, SB RAS, Novosibirsk, 630090, Russia}
\affiliation{Novosibirsk State University, Novosibirsk, 630090, Russia}
\author{A.~P.~Kryukov}
\affiliation{Budker Institute of Nuclear Physics, SB RAS, Novosibirsk, 630090, Russia}
\author{N.~A.~Melnikova}
\affiliation{Budker Institute of Nuclear Physics, SB RAS, Novosibirsk, 630090, Russia}
\affiliation{Novosibirsk State University, Novosibirsk, 630090, Russia}
\author{N.~Yu.~Muchnoy}
\affiliation{Budker Institute of Nuclear Physics, SB RAS, Novosibirsk, 630090, Russia}
\affiliation{Novosibirsk State University, Novosibirsk, 630090, Russia}
\author{A.~E.~Obrazovsky}
\affiliation{Budker Institute of Nuclear Physics, SB RAS, Novosibirsk, 630090, Russia}
\author{E.~V.~Pakhtusova}
\affiliation{Budker Institute of Nuclear Physics, SB RAS, Novosibirsk, 630090, Russia}
\author{E.~A.~Perevedentsev}
\affiliation{Budker Institute of Nuclear Physics, SB RAS, Novosibirsk, 630090, Russia}
\affiliation{Novosibirsk State University, Novosibirsk, 630090, Russia}
\author{K.~V.~Pugachev}
\affiliation{Budker Institute of Nuclear Physics, SB RAS, Novosibirsk, 630090, Russia}
\affiliation{Novosibirsk State University, Novosibirsk, 630090, Russia}
\author{Yu.~A.~Rogovsky}
\affiliation{Budker Institute of Nuclear Physics, SB RAS, Novosibirsk, 630090, Russia}
\affiliation{Novosibirsk State University, Novosibirsk, 630090, Russia}
\author{S.~I.~Serednyakov}
\affiliation{Budker Institute of Nuclear Physics, SB RAS, Novosibirsk, 630090, Russia}
\affiliation{Novosibirsk State University, Novosibirsk, 630090, Russia}
\author{Yu.~M.~Shatunov}
\affiliation{Budker Institute of Nuclear Physics, SB RAS, Novosibirsk, 630090, Russia}
\affiliation{Novosibirsk State University, Novosibirsk, 630090, Russia}
\author{D.~A.~Shtol}
\affiliation{Budker Institute of Nuclear Physics, SB RAS, Novosibirsk, 630090, Russia}
\author{Z.~K.~Silagadze}
\affiliation{Budker Institute of Nuclear Physics, SB RAS, Novosibirsk, 630090, Russia}
\affiliation{Novosibirsk State University, Novosibirsk, 630090, Russia}
\author{I.~K.~Surin}
\affiliation{Budker Institute of Nuclear Physics, SB RAS, Novosibirsk, 630090, Russia}
\author{M.~V.~Timoshenko}
\affiliation{Budker Institute of Nuclear Physics, SB RAS, Novosibirsk, 630090, Russia}
\author{Yu.~V.~Usov}
\affiliation{Budker Institute of Nuclear Physics, SB RAS, Novosibirsk, 630090, Russia}
\author{I.~M.~Zemlyansky}
\affiliation{Budker Institute of Nuclear Physics, SB RAS, Novosibirsk, 630090, Russia}
\affiliation{Novosibirsk State University, Novosibirsk, 630090, Russia}
\author{V.~N.~Zhabin}
\email{V.N.Zhabin@inp.nsk.su}
\affiliation{Budker Institute of Nuclear Physics, SB RAS, Novosibirsk, 630090, Russia}
\affiliation{Novosibirsk State University, Novosibirsk, 630090, Russia}
\author{V.~V.~Zhulanov}
\affiliation{Budker Institute of Nuclear Physics, SB RAS, Novosibirsk, 630090, Russia}
\affiliation{Novosibirsk State University, Novosibirsk, 630090, Russia}

\collaboration{SND Collaboration}

\begin{abstract}
The process \( e^+e^- \to \omega\pi^0 \to \pi^+\pi^-\pi^0\pi^0 \) is studied in the center-of-mass energy region 1.05--2.00 GeV using data with an integral luminosity of about 35~pb$^{-1}$ collected with the SND detector at the VEPP-2000 $e^+e^-$ collider.
In the energy range under study, the value of the measured Born cross section varies from 0.7 to 18 nb.
The statistical uncertainty of the cross section is 2--23\%, while the systematic uncertainty is in the range of 3.0--14.2\%.
The results are consistent with previous measurements but have better accuracy.
\end{abstract}

\maketitle

\section{Introduction}

\begin{figure*}
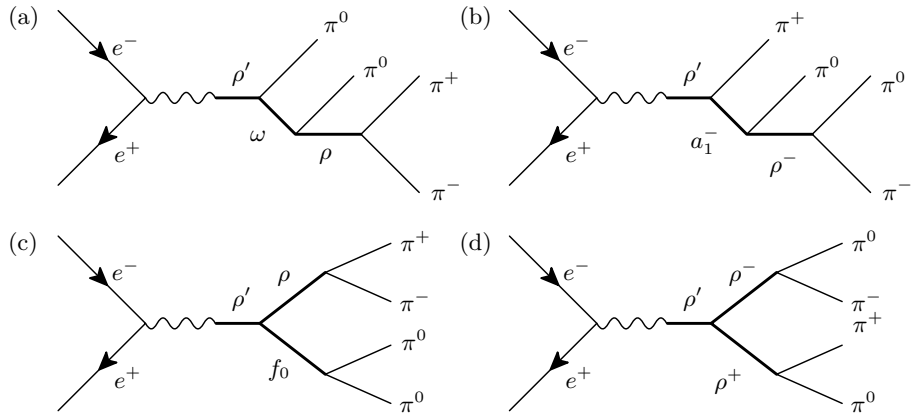

\centering
\setlength{\unitlength}{0.75cm}
\begin{picture}(16,8)
\put(0,7.5){(a)}
\def\svgwidth{8\unitlength}
\put(0.5,4){\import{figs}{diagWpi0_.tex}}
\put(8,7.5){(b)}
\def\svgwidth{8\unitlength}
\put(8.5,4){\import{figs}{diagA1pi_.tex}}
\put(0,3.5){(c)}
\def\svgwidth{8\unitlength}
\put(0.5,0){\import{figs}{diagf0rho_.tex}}
\put(8,3.5){(d)}
\def\svgwidth{8\unitlength}
\put(8.5,0){\import{figs}{diagrhorho_.tex}}
\end{picture}
\caption{
	The Feynman diagrams for the main intermediate states contributing to the process \( e^+e^- \to \pi^+\pi^-\pi^0\pi^0 \): (a)~$ \omega(782)\pi^0 $, (b)~$ a_1 (1260)\pi $, (c)~$ f_0 (980)\rho $, (d)~$ \rho^+ \rho^- $.
	The symbol $\rho^\prime$ denotes a resonance of the $\rho$ family.
}
\label{fig:diag4pi}
\end{figure*}

The process \( e^+e^- \to \pi^+\pi^-\pi^0\pi^0 \) dominates in the hadronic cross section in the center-of-mass (c.~m.) energy region $E \equiv \sqrt{s}$ from 1.2 to 2 GeV and gives a contribution to the hadronic vacuum polarization significant for the calculation of the muon anomalous magnetic moment $(g-2)_\mu$~\cite{davier2019}.
In this energy region, the process \( e^+e^- \to \pi^+\pi^-\pi^0\pi^0 \) has four main intermediate states: $\omega(782)\pi^0$, $a_1 (1260)\pi$, $f_0(980)\rho$, and $\rho^+\rho^-$~\cite{babar17}, the diagrams for which are shown in Fig.~\ref{fig:diag4pi}.
In the region of 1--1.5 GeV, the largest contribution to the $e^+e^- \to \pi^+\pi^-\pi^0\pi^0$ cross section comes from the $\omega(782)\pi^0$ mechanism.
A characteristic feature of this mechanism is a narrow peak near the $\omega$-meson mass in the $\pi^+\pi^-\pi^0$ invariant mass spectrum.
Therefore, it is easily separated both from other mechanisms and from background processes.

This work presents the measurement of the \( e^+e^- \to \omega(782)\pi^0 \to \pi^+\pi^-\pi^0\pi^0 \) cross section in the SND experiment at the VEPP-2000 $e ^+e^-$ collider.
We study in detail all possible sources of systematic uncertainties in the cross section measurement for the $\pi^+\pi^-\pi^0\pi^0$ final state.
The results of this study will be used in the future for precision measurement of the total cross section of the process \(e^+e^- \to\pi^+\pi^-\pi^0\pi^0 \) including all its intermediate states.

The \( e^+e^- \to \omega\pi^0 \) cross section below 2~GeV is saturated with the contributions of the isovector resonances $\rho\equiv\rho(770)$, $\rho^\prime\equiv\rho(1450)$ and $\rho^{\prime\prime}\equiv\rho(1700)$.
The parameters of these resonances are extracted from the fit to the cross section energy dependence with the vector meson dominance (VMD) model~\cite{snd2016}.
This dependence can be also used to predict the hadronic spectrum for the decay \( \tau \to \omega\pi\nu_\tau \)~\cite{CLEO} and its branching fraction~\cite{snd2016}, and, therefore, to test the vector current conservation hypothesis with high precision.

The cross section of the process \( e^+e^- \to \omega\pi^0 \) was measured independently in two channels: \( \omega \to \pi^+\pi^-\pi^0 \) and \( \omega \to \pi^0\gamma \).
Measurements in the \( \omega \to \pi^0\gamma \) channel were carried out with the ND~\cite{nd1986}, SND~\cite{snd2000b,snd2000a}, and CMD-2~\cite{cmd03} detectors at the VEPP-2M collider at c.~m. energies below 1.4 GeV, at the KLOE~\cite{kloe08} experiment near the $\phi$-meson resonance, and with the SND~\cite{snd2016} detector at VEPP-2000 below 2~GeV.
The measurements in the \( \omega \to \pi^+\pi^-\pi^0 \) channel were carried out at the DM2~\cite{dm2} experiment in the energy range 1.35--2.4 GeV, at the VEPP-2M collider with CMD-2~\cite{cmd99} and SND~\cite{snd2000b,snd2003,snd2009} detectors below 1.4 GeV, at the KLOE~\cite{kloe08} experiment near $\phi$, at the BESIII~\cite{bes2021} experiment in the range 2.00--3.08 GeV, and at the BABAR~\cite{babar17} experiment using the initial state radiation (ISR) method in the range 0.92--2.50 GeV.

\section{Experiment}
\label{sec2}

\begin{figure*}
\centering
\includegraphics[width=0.8\textwidth]{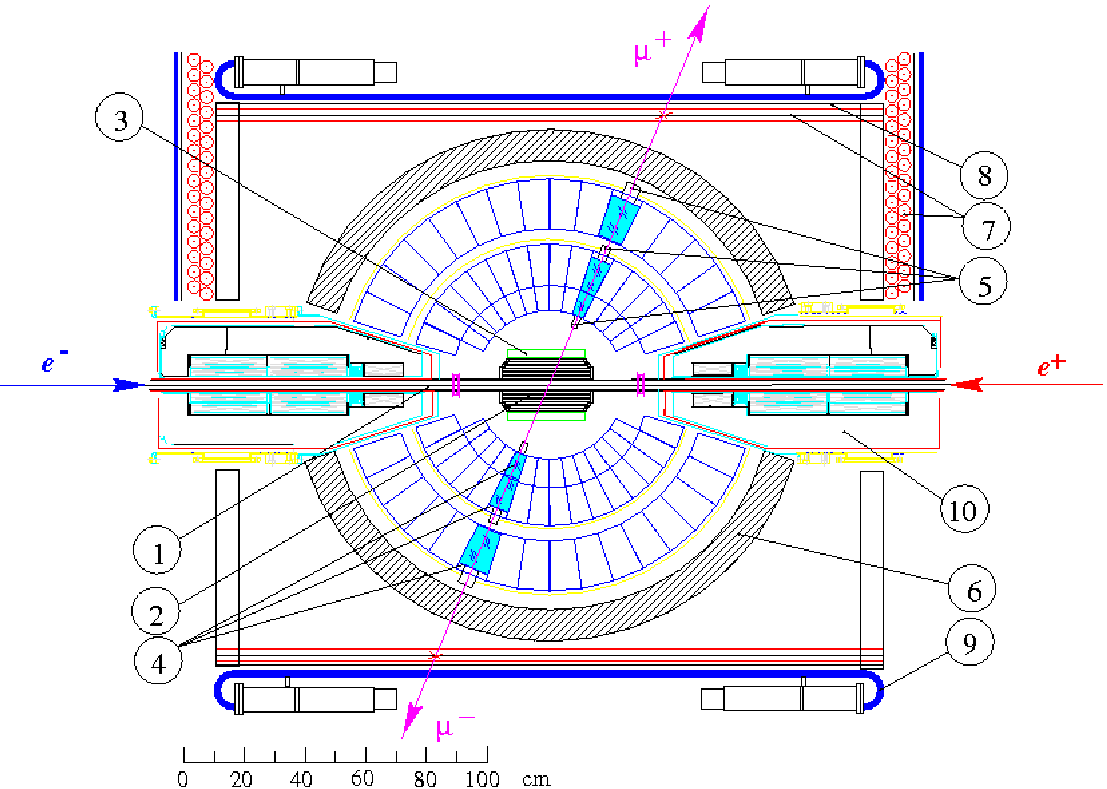}
\caption{
	The schematic view of the SND detector: beam pipe (1), drift chamber (2), aerogel Cherenkov counters (3), NaI(Tl) crystals (4), phototriodes (5), iron absorber (6), muon proportional tubes (7), iron plates (8), muon scintillation counters (9), focusing solenoids of the VEPP-2000 collider (10).
}
\label{fig:sndscheme}
\end{figure*}

SND is a general-purpose nonmagnetic detector~\cite{sndover1,sndover2,sndover3,sndover4} (Fig.~\ref{fig:sndscheme}).
Since 2010, SND has been collecting data at the VEPP-2000 electron-positron collider operating in the 0.3--2 GeV energy range.
Another detector, CMD-3~\cite{cmdover}, is also installed at the collider and collects data simultaneously with SND.

The SND detector consists of the electromagnetic calorimeter, the drift chamber, Cherenkov counters, and the muon system.
The calorimeter consists of 1640 NaI(Tl) crystals arranged in three spherical layers with a total thickness of 13.4 radiation lengths and covers 95\% of the total solid angle (from $18^\circ$ to $162^\circ$ in polar angle).
The nine-layer drift chamber used to detect tracks of charged particles covers 94\% of the total solid angle and has a resolution of $0.45^\circ$ for azimuthal angle and $0.8^\circ$ for the polar angle.
Aerogel Cherenkov counters are used for $\pi-K$ separation.
The muon system is located around the calorimeter and is separated from it by the iron absorber.
It consists of proportional tubes and scintillation counters separated by iron plates and is used, in particular, to suppress the cosmic-ray background.

The analysis is based on data collected by SND in 2011 and 2012 by scanning the c.~m. energy region from 1.05 to 2 GeV with a step of about 25 MeV.
Data with a total integrated luminosity of 34.5 $\text{pb}^{-1}$ were recorded at 53 energy points.
The beam energy $E_b$ was controlled by magnetic field measurements in the collider bending magnets.
In 2012, it was also measured at several points using the back-scattering-laser-light system~\cite{sndcompt,sndcompt2}.
The CMD-3 detector, by measuring the momenta of the final particles in the reactions \( e^+e^- \to e^+e^- \) and \( e^+e^- \to p\bar{p} \), obtained corrections to the energy~\cite{cmdenergy}, which are used in this work.
The accuracy of the c.~m. energy determination is 6 MeV and 2 MeV for the 2011 and 2012 data sets, respectively.

To simulate the process under study, an event generator based on Ref.~\cite{czyz} is used.
It can simulate the intermediate states $\omega(782)\pi^0$, $a_1 (1260)\pi$, $f_0(980)\rho$, and $\rho^+\rho^-$, separately, and in any combinations taking into account the interference between their amplitudes.
The generator includes the emission of additional photons from the initial state~\cite{kuraev,bonneau}.
To calculate the spectrum of the ISR photons, it is necessary to know the Born cross-section energy dependence. For the process under study, it is determined iteratively using data.

The detector response is simulated using the GEANT4 framework~\cite{geant4}.
The simulation takes into account spurious photons and charged tracks arising from superimposing beam-induced background on the events of interest.
To do this, background events recorded during the experiment with a special random trigger are mixed with the simulated events of the process under study as well as physical background processes.

The luminosity is determined using the events of the elastic scattering \(e^+e^- \to e^+e^- \) with a systematic uncertainty of 2\%~\cite{snd_etapp_2015}.

\section{Event selection}

\begin{figure}
\centering
\includegraphics[width=\columnwidth]{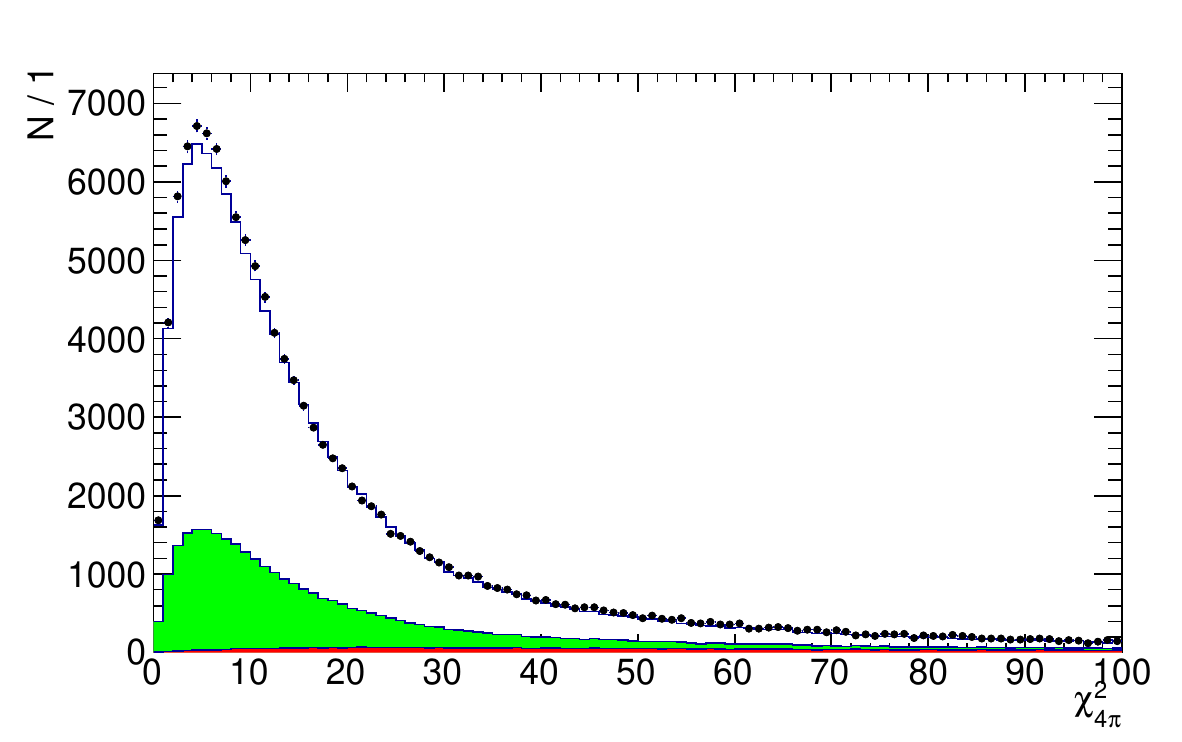}
\caption{
	The distribution of $\chi^2$ of the kinematic fit in the hypothesis \( e^+e^- \to \pi^+\pi^-\pi^0\pi^0 \) for selected experimental events (points with error bars).
	The solid histogram is the sum of simulated distributions for signal and background.
	The distributions are normalized basing on the results of the fit to the $m_{3\pi}$ spectrum.
	The green shaded histogram represents the distribution for the process \( e^+e^- \to \pi^+\pi^-\pi^0\pi^0 \) with intermediate states other than $\omega\pi^0$.
	The red shaded histogram shows the expected contribution from other background processes.
}
\label{fig:chi2mix4pi}
\end{figure}

Events with two charged particles coming from the beam interaction region and at least four photons with energies above 25 MeV are selected for analysis.
Pairs of photons with an invariant mass in the range of 70--200 MeV are considered as $\pi^0$ candidates.
An event must have at least two such candidates.
At energies above 1.8 GeV, to suppress the background from multiphoton processes such as \( e^+e^- \to \pi^+\pi^-3\pi^0 \) and \( \pi^+\pi^-4\pi^0 \), the number of photons $n_\gamma$ in an event is required to be less than six.
At energies below 1.1 GeV, the background from the process \( e^+e^- \to K^+K^- \) is suppressed by the condition on the angle between the directions of charged particles \( \Delta\Psi < 160^\circ \).

Selected events are then kinematically fitted to the hypothesis \( e^+e^- \to \pi^+\pi^-\pi^0\pi^0 \) with six constraints: four conditions of total energy and momentum balance and two conditions that the invariant masses of photon pairs are equal to the $\pi^0$ mass.
If there are several photon combinations with two $\pi^0$ candidates in an event, then the combination with the best $\chi^2$ of the kinematic fit ($\chi^2_{4\pi}$) is chosen.
The $\chi^2_{4\pi}$ distribution for selected data and simulated events is shown in Fig.~\ref{fig:chi2mix4pi}.
The condition $\chi^2_{4\pi} < 40$ is imposed.

For each event that passed the selection conditions, the invariant masses for two $\pi^+\pi^-\pi^0$ combinations are calculated, and the mass value ($m_{3\pi}$) closest to the $\omega$ mass is chosen.
Events with \( 650 < m_{3\pi} < 900 \) MeV are selected for further analysis.

\section{Background processes}
\label{sec:background}

\begin{figure*}
\centering
\includegraphics[width=0.8\textwidth]{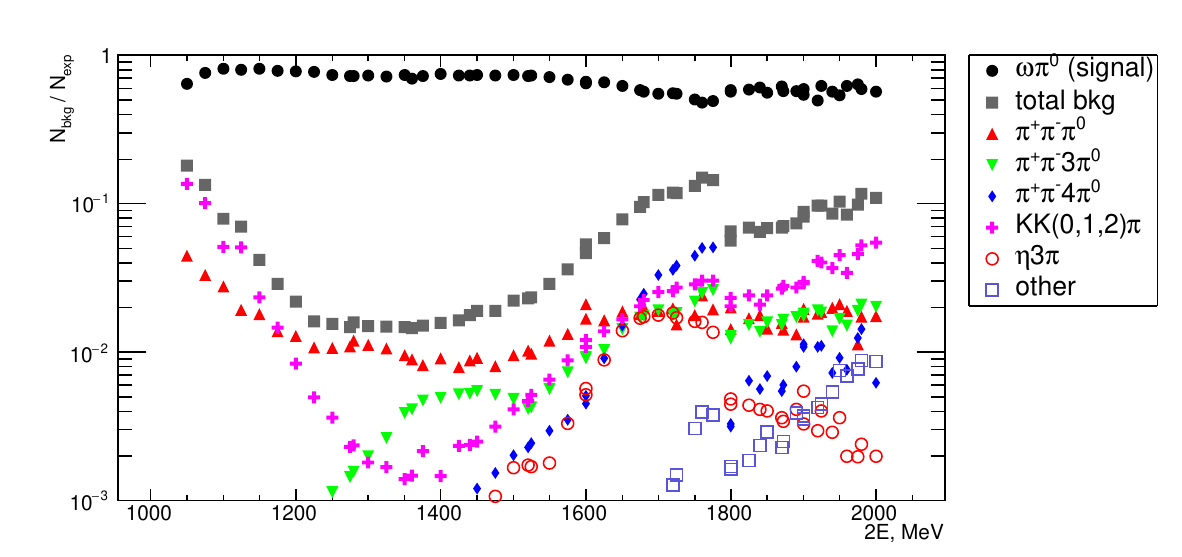}
\caption{
	The relative contribution of the processes \( e^+e^- \to \omega\pi^0 \), $\pi^+\pi^-\pi^0$, $K \bar{K}$, $\pi^+\pi^- 3\pi^0$, $\pi^+\pi^- 4\pi^0$, and sum of other background processes other than $\pi^+\pi^-\pi^0\pi^0$ to the number of selected events calculated using simulation, taking into account the scale factors obtained from data.
}
\label{fig:cross_mix}
\end{figure*}

The contribution of background processes other than \( e^+e^- \to \pi^+\pi^-\mbb\pi^0\pi^0 \) is calculated using simulation.
The following processes with two charged particles are studied:
\( e^+e^- \to \pi^+\pi^-\pi^0 \), $\pi^+\pi^- 3\pi^0$, $\pi^+\pi^- 4\pi^0$,
$\pi^+\pi^-\eta$, $\pi^+\pi^-\pi^0\eta$, $\pi^+\pi^-\mbb2\pi^0\eta$,
$K \bar{K}$, $K \bar{K} \pi$, $K^+K^-\eta$ and $K^+K^-\pi^0\pi^0$.
In events of the processes with less than four photons in the final state, additional photons originate from ISR, beam-induced background, nuclear interaction of charged pions/kaons in the calorimeter, kaon decays, and splitting of electromagnetic showers.
Processes with four charged particles were also studied, but their contribution is found to be negligible.

The number of events of each background process at each energy point is calculated as
\[ N_{bkg} = \sigma_{bkg} \varepsilon_{bkg} L, \]
where $\sigma_{bkg}$ is the visible cross section of the background process (see Eq.~(\ref{eq:radcor})) calculated using existing data on the Born cross section,
$\varepsilon_{bkg}$ is its detection efficiency for the \( e^+e^- \to \omega\pi^0 \) selection conditions determined using simulation,
$L$ is the integrated luminosity collected at a given energy point.

The largest background contribution at \( E < 1.2 \) GeV comes from the processes \( e^+e^- \to K\bar{K} \), which have a large cross section due to the proximity of the $\phi(1020)$ resonance.
To suppress the background from the process \( e^+e^- \to K^+K^- \) at \( E < 1.1 \) GeV, a special condition is applied to the angle between the directions of charged particles \( \Delta\Psi < 160^\circ\).

In the region \( 1.2 < E < 1.7 \) GeV, the main background process is \( e^+e^- \to \pi^+\pi^-\pi^0(\gamma) \), usually with an additional photon emitted from the initial state.
The Born cross section for the process \( e^+e^- \to \pi^+\pi^-\pi^0 \) is taken from Refs.~\cite{snd_3pi_1,snd_3pi_2,snd_3pi_3}, and for processes \( e^+e^- \to K\bar{K} \) from Refs.~\cite{pdg,snd_kk,babar_kk}.

Above 1.7 GeV, the processes \( e^+e^- \to \pi^+\pi^-3\pi^0 \) and \( e^+e^- \to \pi^+\pi^-4\pi^0 \) give the main contribution to background events.
To suppress events of the latter process at energies above 1.8 GeV, the restriction on the number of photons \( n_\gamma < 6 \) is used.
After imposing this condition, the process \( e^+e^- \to K^+K^-\pi^0\pi^0 \) becomes the dominant background source.
The Born cross sections for these processes are taken from Refs.~\cite{babar_2pi3pi0,babar_2pi4pi0,babar_4pic1pi0,babar_kk2pic}.
For the background process \( e^+e^- \to \pi^+\pi^-3\pi^0 \), the intermediate states $\pi^+\pi^-\eta$ and $\omega\pi^0 \pi^0$ are simulated separately.
As shown in Ref.~\cite{babar_2pi3pi0}, these two contributions saturate the \( e^+e^- \to \pi^+\pi^-3\pi^0 \) cross section below 1.8 GeV.
For the rest of the \( e^+e^- \to \pi^+\pi^-3\pi^0 \) cross section, the uniform pion phase space distribution is used.
In the process \( e^+e^- \to \pi^+\pi^ - 4\pi^0 \), with the exception of the $\omega\eta$ intermediate state, the pions are also generated uniformly over phase space.

The contributions of three background processes: \( e^+e^- \to \pi^+\pi^-\pi^0 \), $\pi^+\pi^-3\pi^0$, and $K^+ K^-$ are estimated individually from data.
To do this, a simultaneous fit to the $m_{3\pi}$ distributions in two classes of events is performed with a sum of signal and background distributions.
The first class includes events with \( \chi^2_{4\pi} < 40 \) passing the condition $\bar{c}$, while the second class includes events satisfying the conditions $c$ and \( \chi^2_{4\pi} < 100 \).
Here $c$ is the condition used to select events of a specific background process: \( \chi^2_{3\pi} < 10 \) for $\pi^+\pi^-\pi^0$, \( \chi^2_{5\pi} < 20 \) for $\pi^+\pi ^-3\pi^0$, and \( \Delta\Psi > 165^\circ \) for $K^+K^-$.
In the case of the $\pi^+\pi^-3\pi^0$ process, the additional conditions \( \chi^2_{6\pi} > 20 \) and \( n_\gamma \leq 6 \) are imposed on events of the both classes to suppress \( e^+e^- \to \pi^+\pi^-4\pi^0 \) background.
The parameters $\chi^2_{3\pi}$, $\chi^2_{5\pi}$ and $\chi^2_{6\pi}$ are $\chi^2$ of the kinematic fits in the hypotheses \( e^+e^- \to \pi^ +\pi^-\pi^0 \), \( \pi^+\pi^-3\pi^0 \), and \( \pi^+\pi^-4\pi^0 \), respectively.
In the fit to the $m_{3\pi}$ distributions, the free parameters are the number of signal events in each class and the scale factor for the background process under study.
The contributions of remaining background processes are fixed at the calculated values.
The first class is needed mainly to determine the ratio between the number of $\omega\pi^0$ events and the number background events from other mechanisms of the \( e^+e^- \to \pi^+\pi^-\pi^0\pi^0 \) process.
The second class, enriched with background events passed the selection conditions as close as possible to the nominal ones, is needed to determine the scale factor for the background process under study.

For the process \( e^+e^- \to \pi^+\pi^-\pi^0 \), the obtained scale factor is $0.85 \pm 0.16$ for $E<1.2$ GeV, $1.96 \pm 0.24$ for $E >1.6$ GeV and varies
linearly between these values at $1.2<E<1.6$ GeV.
For \( e^+e^- \to K^+K^- \), it is $1.50 \pm 0.04$ for $E<1.12$ GeV and $1.00 \pm 0.14$ above.
For \( e^+e^- \to \pi^+\pi^-3\pi^0 \), the scale factor is $0.30 \pm 0.03$ over the entire energy range.

A similar procedure is used to estimate the total contribution of other background processes.
In this case, the first and second classes include events with \( \chi^2_{4\pi} < 40 \) and \( 60 < \chi^2_{4\pi} < 100 \), respectively, while the process specific conditions are not used.
For the three processes discussed above, the found scale factors are applied.
The scale factor $1.2 \pm 0.3$ found for the entire energy range is consistent with unity.
Low statistical accuracy does not allow to study its energy dependence.
Therefore, in further analysis, it is fixed equal to unity with a systematic uncertainty of 50\%.

The relative contribution of different background processes to the number of selected events, calculated by simulation and corrected by the scale factors defined above, is shown in Fig.~\ref{fig:cross_mix}.

\section{Determining the number of \( e^+e^- \to \omega\pi^0 \) events}
\label{sec:omegafitting}

\begin{figure*}
\centering
\includegraphics[width=0.45\textwidth]{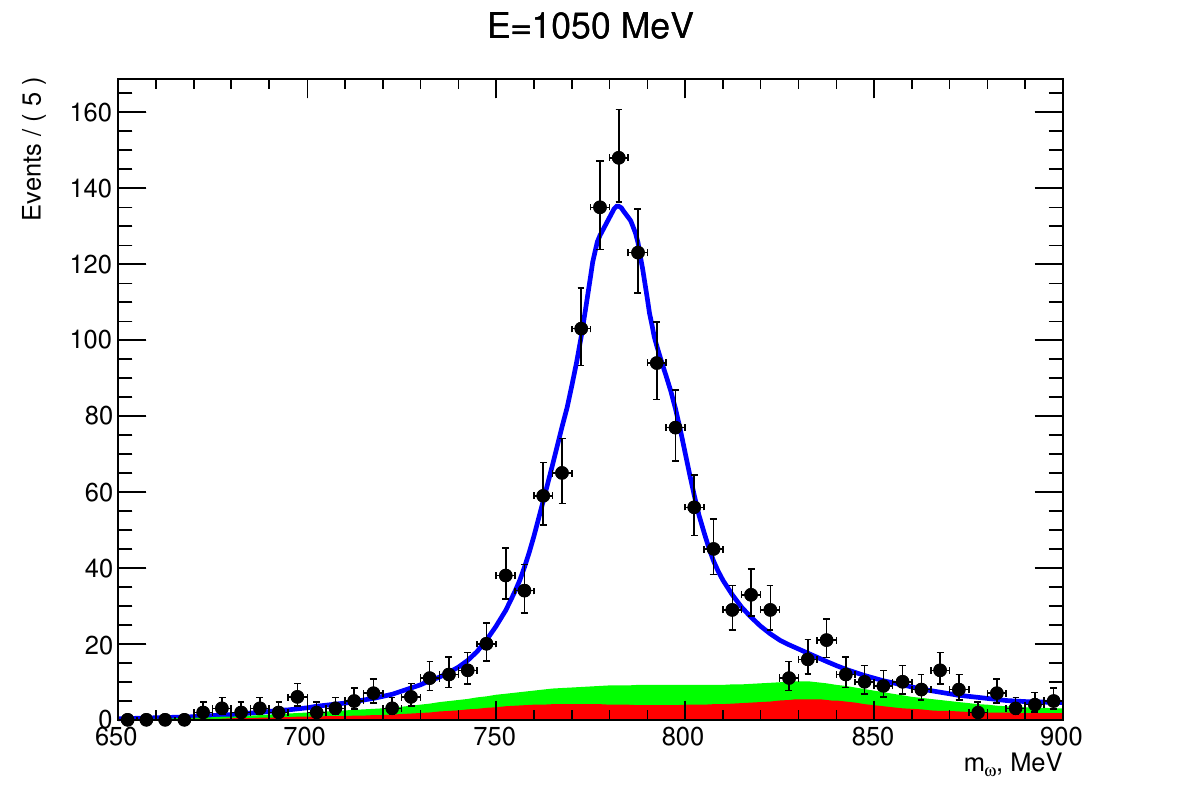}
\includegraphics[width=0.45\textwidth]{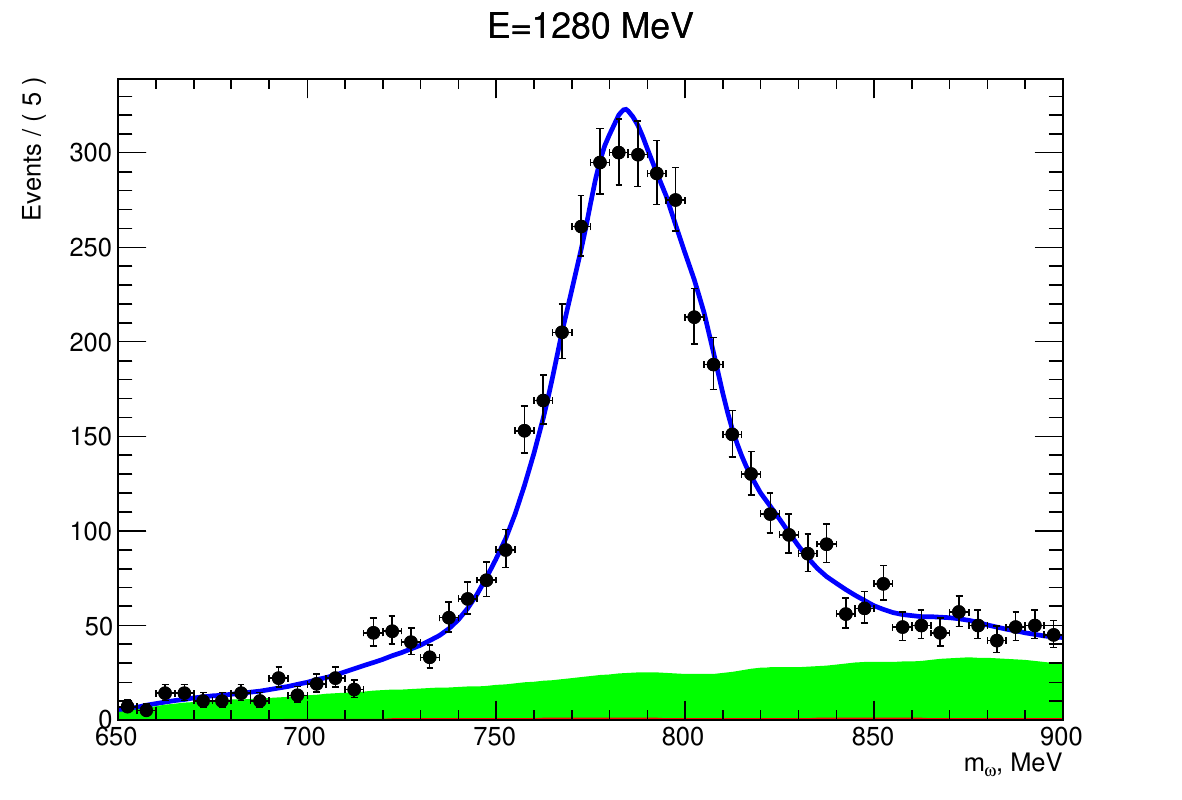}
\includegraphics[width=0.45\textwidth]{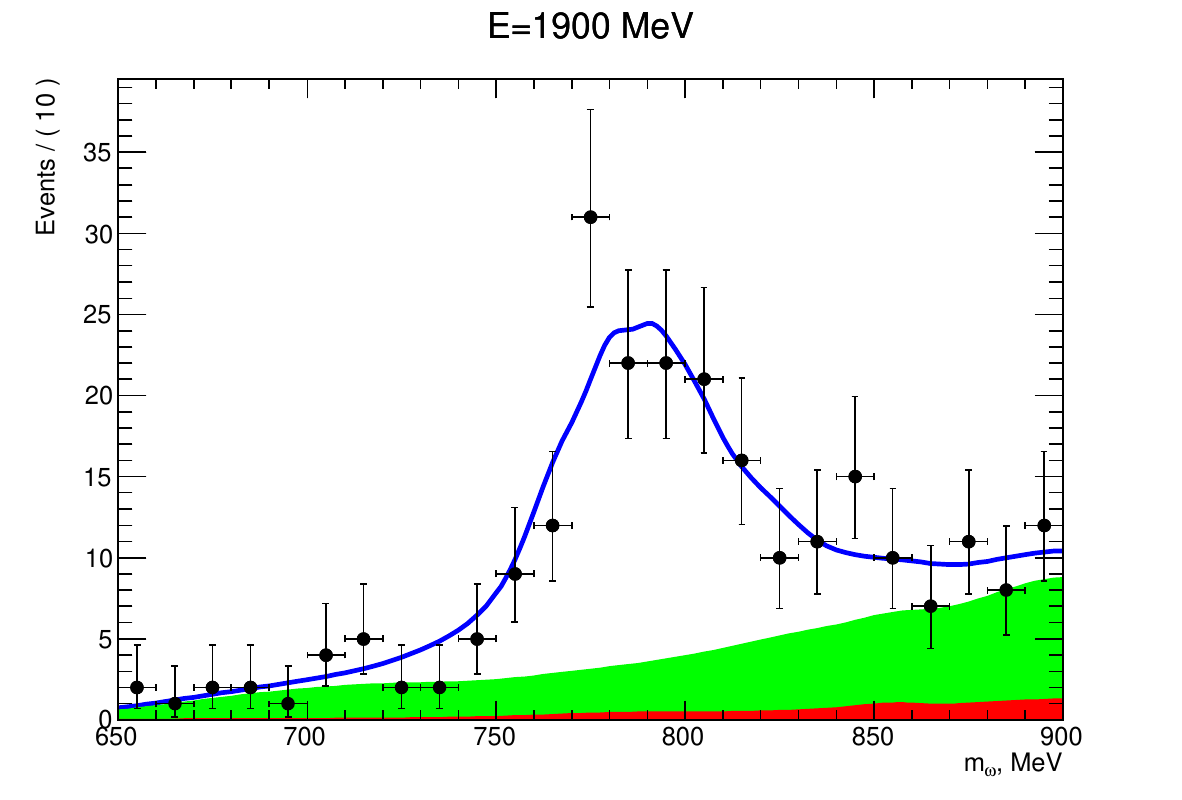}
\includegraphics[width=0.45\textwidth]{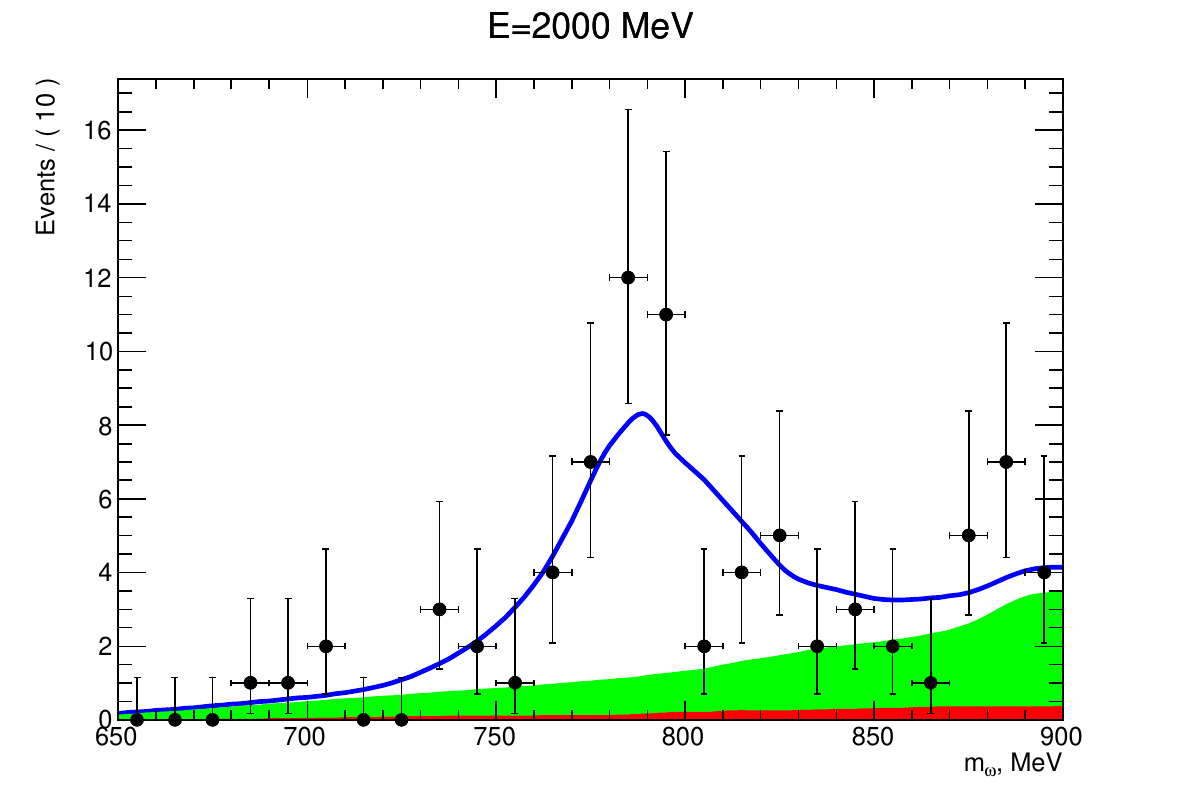}
\caption{
	The distribution of the $\pi^+\pi^-\pi^0$ invariant mass closest to the $\omega$ mass for selected data events with $E = 1050$, 1280, 1900, 2000 MeV (points with error bars).
	The solid curve is the result of the fit to the data distribution with a sum of signal and background distributions described in the text.
	The green dashed region is the fitted contribution from background $\pi^+\pi^-\pi^0\pi^0$ events with intermediate states other than $\omega\pi^0$.
	The red dashed region is the expected background from other processes.
}
\label{fig:massfit}
\end{figure*}

The number of $\omega\pi^0$ events at each energy point is determined from the fit to the $m_{3\pi}$ distribution with a sum of signal and background distributions.
The $m_{3\pi}$ distributions for selected experimental events in four energy points are shown in Fig.~\ref{fig:massfit}.

The fit is performed by an unbinned maximum likelihood method using RooFit~\cite{roofit}.
The shape of signal and background distributions is determined from simulation using the kernel estimation technique~\cite{kernel}.
The contributions of the intermediate states $\omega\pi^0$, $\rho^+\rho^-$, $a_1\pi$, and $f_0\rho$ ($N_{\omega\pi^0}$, $N_{ \rho^+\rho^-}$, $N_{a_1\pi}$, and $N_{f_0\rho}$) are free fit parameters, while the contributions from background processes other than $\pi^+\pi^-\pi^0\pi^0$ are fixed at the calculated values, as described in Sec.~\ref{sec:background}.
The numbers of \( e^+e^- \to \omega\pi^0 \) events ($N_{\omega\pi}$) obtained from the fit to the $m_{3\pi}$ distributions at different energy points are listed in Table~\ref{tab:crossfit} with their statistical uncertainties.

\begin{figure}
\centering
\includegraphics[width=0.45\textwidth]{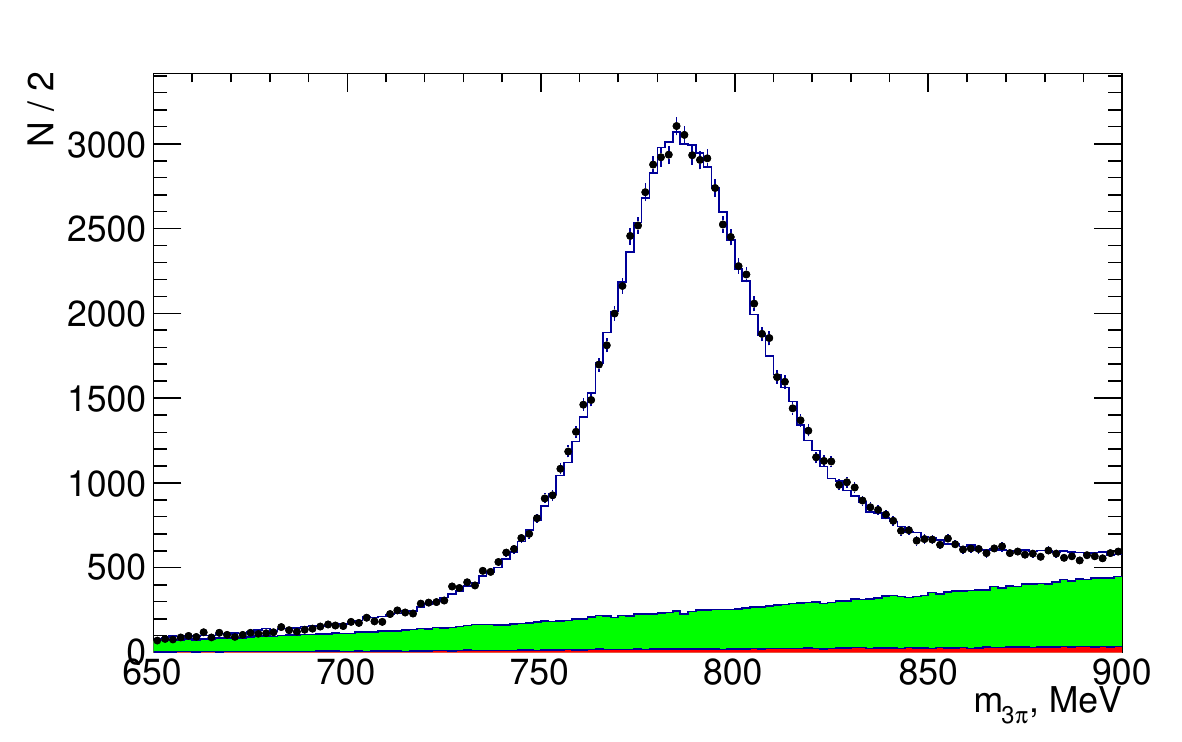}
\caption{
	The fit to the $m_{3\pi}$ distribution for all energy points taking into account the mass shift and smearing of the simulated signal distribution.
}
\label{fig:masshist}
\end{figure}

Incorrect simulation of the angular and energy distributions of detected particles can lead to a difference between the $m_{3\pi}$ distributions in data and simulation.
This is especially significant for the relatively narrow signal distribution.
The difference is parametrized by the shift ($\Delta m$) of the simulated signal distribution and its Gaussian smearing ($\sigma_{\rm G}$).
To determine these parameters, we fit to the $m_{3\pi}$ distribution for all energy points shown in Fig.~\ref{fig:masshist}.
It is found that the best agreement between data and simulation is achieved when the simulation is shifted by \( \Delta m=1.4 \pm 1.0 \) MeV to the right and smeared with \( \sigma_{\rm G} = 5.5 \pm 2.0 \) MeV.
The introduction of $\Delta m$ and $\sigma_{\rm G}$ changes the fitted number of $\omega\pi^0$ events by 1\%.

It should be noted that the used parametrization of the background from $\pi^+\pi^-\pi^0\pi^0$ events by the sum of distributions for three intermediate states does not take into account the interference between their amplitudes.
This means that the obtained numbers of events for mechanisms $a_1\pi$, $f_0\rho$ and $\rho^+\rho^-$ have no physical meaning.
We assume that such a parametrization provides enough freedom to describe the background spectrum observed in data and correctly determine the total number of background events.
To estimate the possible systematic uncertainty associated with the description of the $\pi^+\pi^-\pi^0\pi^0$ background shape, an alternative fit is performed in the model with only one background mechanism $a_1\pi$.
This mechanism dominates in the $\pi^+\pi^-\pi^0\pi^0$ background at energies below 1.5 GeV~\cite{kozyrev} and remains significant at higher energies.
The difference between the two parametrizations of the $\pi^+\pi^-\pi^0\pi^0$ background in the number of fitted signal events does not exceed 0.4\% at $E<1.5$ GeV and reaches 3.9\% at 2 GeV.
This difference is taken as an estimate of the model uncertainty in the number of $\omega\pi^0$ events.

The systematic uncertainty arising from the subtraction of background from processes other than \( e^+e^- \to \pi^+\pi^-\pi^0\pi^0 \) is estimated by varying the contribution of each background process by one standard deviation.
For the $e^+e^-\to 5\pi$ ($6\pi$) process, the model uncertainty is also taken into account, which is estimated from the difference in detection efficiency between simulations using the phase space model and the $\omega\pi^0\pi^0$ ($\omega\eta$) model.
The total systematic uncertainty due to this background subtraction is less than 0.6\%.

Another source of systematic uncertainty in the number of signal events is the interference between $\omega\pi^0$ and other mechanisms of the \( e^+e^- \to \pi^+\pi^-\pi^0\pi^0 \) process.
Due to the finite resolution of the $\pi^+\pi^-\pi^0$ mass measurement, all sign-alternating interference effects leading to distortion of the $\omega$-meson line shape are not visible in data distributions.
The interference only leads to an increase or decrease in the number of signal events determined from the fit relative to the true one.
The interference effect is studied with the Monte Carlo event generator mentioned in Sec.~\ref{sec2}, which includes the four intermediate states $\omega\pi^0$, $\rho^+\rho^-$, $a_1\pi$, and $f_0\rho$.
We calculate the difference in cross section between simulations with ($\sigma_1+\sigma_2+\sigma_{\rm int}$) and without ($\sigma_1+\sigma_2$) interference for each background intermediate state paired with the $\omega\pi^0$ state and obtain the overlap integral $\sigma_{\rm int}/\sqrt{\sigma_1\sigma_2}$.
The Monte-Carlo integration is performed over a phase space restricted by the condition \( 760 < m_{3\pi}^\text{truth} < 820 \) MeV.
Using the overlap integrals, the background-to-signal ratio determined from the fit to the data $m_{3\pi}$ spectrum, and the information about the $f_0\rho$ relative fraction from Ref.~\cite{babar17}, we can calculate the shift in the number of $\omega\pi^0$ events due to the interference.
The values of the overlap integrals depend on the relative phases between the amplitudes.
We vary them from 0 to $2\pi$ to obtain the maximum deviation in the number of signal events.
This deviation is taken as an estimate of the systematic uncertainty associated with the interference of the $\omega\pi^0$ mechanism with other mechanisms contributing to the $\pi^+\pi^-\pi^0\pi^0$ final state.
It does not exceed 0.5\% at 1.05 GeV and 13\% at 2 GeV.
The relatively small value of the uncertainty is due to the narrowness of the $\omega$ resonance.

\section{Detection efficiency}
\label{sec:effcorr}

\begin{figure}
\centering
\includegraphics[width=0.47\textwidth]{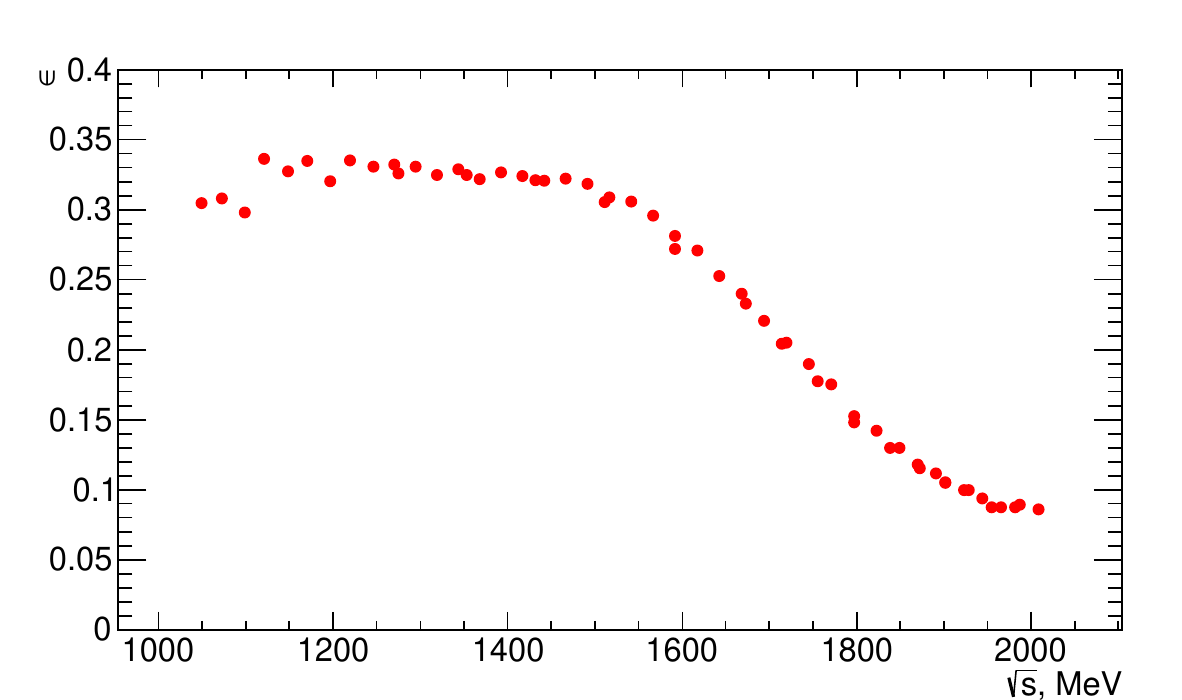}
\caption{
	Detection efficiency of the process \( e^+ e^- \to \omega\pi^0 \to \pi^+ \pi^-\pi^0\pi^0 \), calculated from the simulation, taking into account the corrections determined using experimental data.
}
\label{fig:efficiency}
\end{figure}

The detection efficiency for the process \( e^+ e^- \to \omega\pi^0 \to \pi^+ \pi^-\pi^0 \pi^0 \) calculated from the simulation is shown in Fig.~\ref{fig:efficiency}.
The decrease in the efficiency above 1.5 GeV is associated with a drop in the \( e^+e^- \to \omega\pi^0 \) Born cross section in this region (see section~\ref{sec:borncross}).
As a result of this drop, the fraction of events with the emission of an energetic photon from the initial state, which are rejected by the condition $\chi^2_{4\pi} < 40$, increases.
Steps in the efficiency at energies of 1.1 and 1.8 GeV arise due to changes in the selection conditions at these points.
The nonstatistical scatter in the values of the detection efficiency between neighboring energy points is caused by the change in the experimental conditions during the data taking, in particular, by the change in the number of bad channels in the calorimeter.

The inaccuracy of simulation of the angular and energy distributions of the reconstructed particles leads to a difference between simulation and data in the detection efficiency.
Therefore, the corrections to the detection efficiency associated with the selection conditions are obtained using data.
They are listed in Table~\ref{tab:effcorres}.
It is seen from Table~\ref{tab:effcorres} that the photon loss correction is calculated together with the correction for the condition $\chi^2_{4\pi}<40$.
This is due to the fact that a fake photon, which arises, for example, due to the nuclear interaction of charged pions with the detector material or the beam-induced background, can be added to the event with a lost photon.
Such an event passes the condition $n_\gamma\geq 4$ and with high, but not 100\% probability will be rejected by the condition on $\chi^2_{4\pi}$.
Since the fake photons are poorly simulated, it is expedient to determine both corrections together.

The corrections are calculated as follows
\begin{equation}
\label{eq:cor}
	( 1 - \delta_{\text{eff}}) =
	\left(\frac{N_1+N_2}{N_1}\right)_{\rm data}\bigg/
	\left(\frac{N_1+N_2}{N_1}\right)_{\rm MC},
\end{equation}
where $N_1$ is the number of $\omega\pi^0$ events selected with the standard condition,
and $N_2$ is the number of $\omega\pi^0$ events that are added after loosening the condition.
The numbers $N_1$ and $N_2$ in data are obtained from the fit to the $m_{3\pi}$ spectrum.

To find the correction for photon loss and the condition \( \chi^2_{4\pi} < 40 \), a special kinematic fit is applied to events with \( n_\gamma \geq 3 \) in the hypothesis \( e^+e^- \to \pi^+\pi^-\pi^0\pi^0 \), which uses the parameters of only three photons.
The parameters of the fourth photon are determined from the fit.
The condition \( \chi^2_{4\pi,\text{ lost }\gamma} < 2 \) is imposed on $\chi^2$ of this fit.
To suppress beam background events, the condition on total energy deposition in the calorimeter \( E_\text{tot} / E_b > 0.75 \) is used.
Then, for one of the two tracks, chosen randomly, the difference between $z$ coordinates of the interaction point and the point at the track closest to the beam axis is required to be \( |z_0| < 7.5 \) cm (the $z$ axis is oriented along the colliding beams).
The number of beam background events satisfying this condition is determined from the fit to the $z_0$ distribution in the range \( |z_0| < 15 \) cm.
The signal events have a normal distribution in $z_0$, while the beam background is uniform with a slight slope.
The shape of the $m_{3\pi}$ distribution for beam background events is determined using events from the regions \( 7.5 < |z_0| < 15 \)~cm, in which the contribution of signal events is negligible.
The number $N_1$ includes events satisfying the standard condition \( \chi^2_{4\pi} < 40 \), and $N_2$ includes events with \( \chi^2_{4\pi} > 40 \) and events with \( n_\gamma = 3 \).
The $m_{3\pi}$ spectra obtained with the two selection conditions are fitted simultaneously.
The free fit parameters are $N_1$, $N_2$, \( N_{\rho^+\rho^-} / N_{\omega\pi^0} \), \( N_{a_1\pi} / N_{\omega\pi^0} \), and \( N_{f_0\rho} /\mbb N_{\omega\pi^0} \).
It is found that the correction calculated using Eq.~(\ref{eq:cor}) is independent of energy.
Its average value is listed in Table~\ref{tab:effcorres}.

\begin{figure}
\centering
\includegraphics[width=\columnwidth]{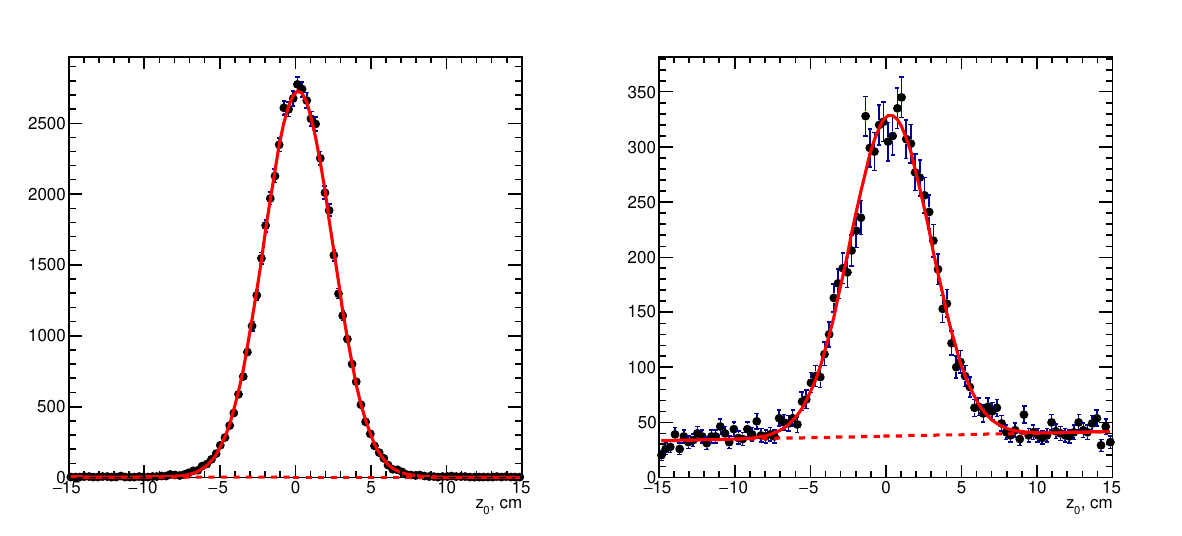}
\caption{
	The $z_0$ distribution for data events (points with error bars) with \( \chi^2_{4\pi,\text{lost }\pi^\pm} < 3 \).
	The curve is the result of the fit described in the text.
	The left (right) panel represents events with two (one) charged tracks.
}
\label{fig:z0_1c}
\end{figure}

To calculate the correction for track loss in the tracking system, a kinematic fit is carried out in the hypothesis \( e^+e^- \to \pi^+\pi^-\pi^0\pi^0 \) using the parameters of one charged particle and four photons.
The fit recovers the parameters of the missing charged particle.
In the case of events with two charged tracks, a candidate for the kinematic fit is
chosen randomly.
The condition \( \chi^2_{4\pi,\text{ lost }\pi^\pm} < 3 \) is imposed on $\chi^2$ of the kinematic fit.
The beam background is suppressed by the condition \( |z_0| < 5 \) cm.
Its residual contribution is determined in the same way as for the photon loss correction.
The results of the fit to the $z_0$ distributions for events with one and two charged tracks are shown in Fig.~\ref{fig:z0_1c}.
The number $N_1$ includes events containing two charged particles, and $N_2$ includes events with only one charged particle.
The correction value determined by the method described above is given in Table~\ref{tab:effcorres}.
This correction also does not depend on the energy.

The corrections for the conditions \( \Delta\Psi < 160^\circ \) and \( n_\gamma < 6 \) are calculated in a similar way.
The obtained values of the total efficiency correction for three energy ranges with different selection conditions are listed in Table~\ref{tab:effcorres}.
The detection efficiency corrected for the difference between data and simulation is listed in Table~\ref{tab:crossfit}.

\begin{table}
\caption{The corrections to the detection efficiency.}
\label{tab:effcorres}
\begin{ruledtabular}
\begin{tabular}{lc}
\multicolumn{1}{c}{Correction} & $1 + \delta_\text{eff}$ \\ \hline
Photon loss and \( \chi^2_{4\pi} < 40 \) & $ -1.4 \pm 2.0 $ \% \\
Track loss & $ -1.1 \pm 0.7 $ \% \\
\( n_\gamma < 6 \) & $ +0.3 \pm 1.0 $ \% \\
\( \Delta\Psi < 160^\circ \) & $ -0.4 \pm 0.1 $ \% \\
\hline
Total at $E < 1.1$ GeV & $ -2.9 \pm 2.1 $ \% \\
Total at $1.1 < E < 1.8$ GeV & $ -2.5 \pm 2.1 $ \% \\
Total at $E > 1.8$ GeV & $ -2.2 \pm 2.3 $ \% \\
\end{tabular}

\end{ruledtabular}
\end{table}

\section{Born cross section}
\label{sec:borncross}

\begin{figure}
\centering
\includegraphics[width=0.5\columnwidth]{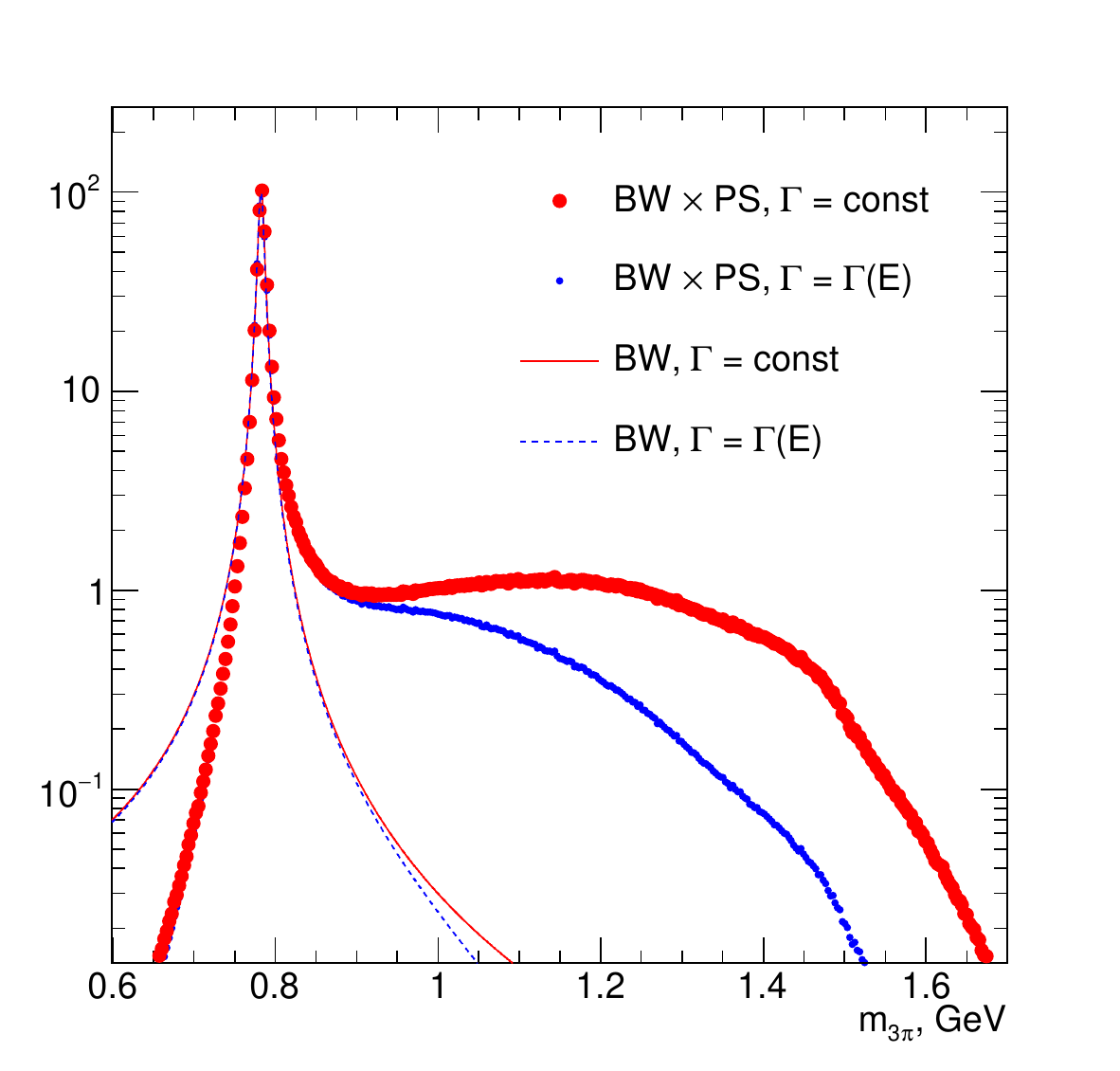}%
\includegraphics[width=0.5\columnwidth]{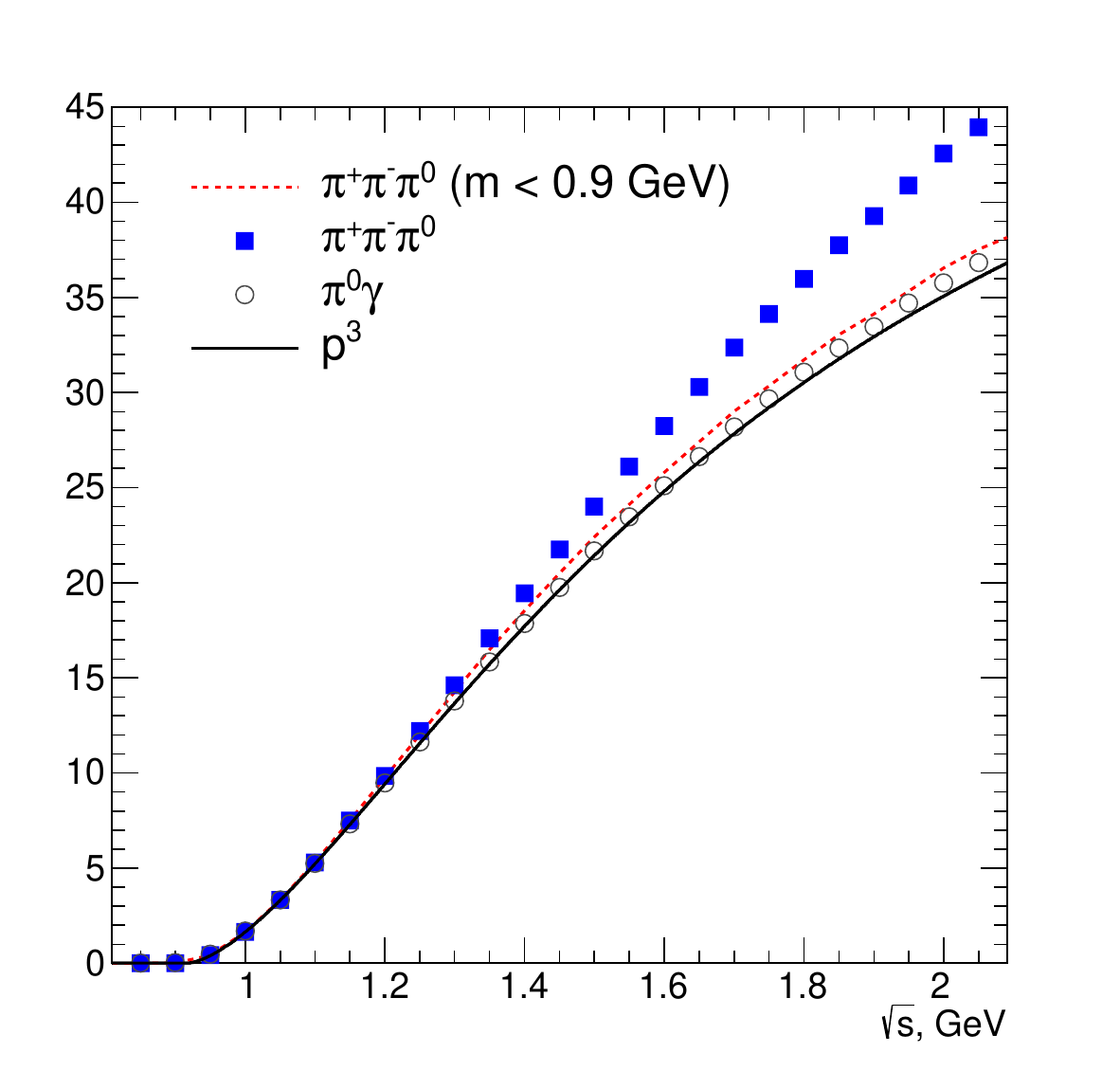}
\caption{
	Left panel: The $\omega$ line shape in the process \( e^+e^- \to \omega\pi^0 \to \pi^+\pi^-\pi^0\pi^0 \) at \( E = 2 \) GeV (BW$\times$PS) compared with the Breit-Wigner line shape without the phase space factor ($BW$) in models with a constant $\omega$-meson width ( $\Gamma =\mathrm{const}$) and with an energy-dependent width ($\Gamma =\Gamma(E)$).
	Right panel: The factor \( 4\pi\alpha^2 g_{\rho\omega\pi}^2 P_f(E) / (E^3 f_\rho^2) \) for the processes \( e^+e^- \to \omega\pi^0 \to \pi^+\pi^-\pi^0\pi^0 \) and \( e^+e^- \to \omega\pi^0 \to \pi^0\pi^0\gamma \) compared with the narrow $\omega$-meson approximation ($p^3$).
}
\label{fig:omlnshape}
\end{figure}

The Born cross section for the process \( e^+e^- \to \omega\pi^0 \to \pi^+\pi^-\pi^0\pi^0 \) can be factorized as~\cite{kardopo13}
\begin{equation}
\label{eq:borncs}
	\sigma_\text{born}(E) = \frac{4\pi\alpha^2}{E^3}
	\left| F_{\gamma\omega\pi}(E) \right|^2 P_f(E),
\end{equation}
where $F_{\gamma\omega\pi}(E)$ is the form factor for the transition $\gamma^*\omega\pi^0$,
$P_f(E)$ is a factor describing the phase space of the final state $\omega\pi^0$.
In the infinitely narrow $\omega$-meson approximation, \( P_f(E) = B(\omega\to\pi^+\pi^-\pi^0)p_{\pi^0}^3/3 \), where $B(\omega\to\pi^+\pi^-\pi^0)$ is the \( \omega \to \pi^+\pi^-\pi^0 \) branching fraction and $p_{\pi^0}$ is the $\pi^0$ momentum.
A more precise expression takes into account the finite $\omega$-resonance width, and the dependence of the $\pi^+\pi^-\pi^0$ phase space on its four-momentum $q^2$ squared~\cite{theta2pi}:
\begin{equation}
	P_f(E) \propto \int \frac
		{ \sqrt{q^2}\, \Gamma_{\omega\to 3\pi}(q^2)\, p_{\pi^0}^3(E,q^2) }
		{ (q^2-m_\omega^2)^2 + q^2 \Gamma_\omega^2(q^2) }\, dq^2,
\label{eq:Pf}
\end{equation}
where $\Gamma_{\omega \to 3\pi}(q^2)$ and $\Gamma_\omega(q^2)$ are the mass-dependent partial and total widths of the $\omega$ meson.
Eq.~(\ref{eq:Pf}) does not take into account the interference between two three-pion combinations.
The calculation taking into account the interference is performed by the Monte Carlo method.
Figure~\ref{fig:omlnshape} (left) shows the mass spectrum of the $\pi^+\pi^-\pi^0$ system compared to the Breit-Wigner formula for the $\omega$ resonance.
The tail in the distribution at large \( q^2 = m_{3\pi}^2 \) is explained by the fast growth of the partial width $\Gamma_{\omega\to 3\pi}$ with increasing $q^2$.
The presence of this tail leads to a significant difference between $P_f(E)$ for $\pi^+\pi^-\pi^0$ and the expression obtained for the narrow $\omega$ meson or the phase space for the decay of $\omega \to \pi^0\gamma$.

To limit too fast growth of the partial widths of the resonances, the $q^2$-dependent Blatt-Weiskopf factors can be introduced, which can strongly modify the theoretical predictions both for the tail in the mass spectrum $\pi^+\pi^-\pi^0$ and $P_f(E)$ dependencies.
At present, the form of these factors are not known exactly.
This introduces a model uncertainty in the measured cross section.
For \( m_{3\pi} > 0.9 \) GeV, it is impossible to experimentally distinguish between $\omega\pi^0$ events and events of other mechanisms of the \( e^+e^- \to \pi^+\pi^-\pi^0\pi^0 \) process.
The number of events $\omega\pi^0$ in this region has to be extrapolated from the region \( m_{3\pi} < 0.9 \)~GeV using a model.

To eliminate this model dependence, the $\omega\pi^0$ cross section can be determined by introducing the constraint \( m_{3\pi} < 0.9 \)~GeV.
This constraint is imposed on both data and simulated events.
In particular, the detection efficiency and the factor $P_f(E)$ are calculated from events with \( m_{3\pi}^\text{truth} < 0.9 \)~GeV.
The factor $P_f(E)$ redefined in this way is shown in Fig.~\ref{fig:omlnshape} (right).
It is seen that its energy dependence is close to that for a narrow resonance.
It should be noted that the value of the form factor $F_{\gamma\omega\pi}(E)$ is independent of the $\omega\pi^0$ cross section definition.

The visible cross section $\sigma_\text{vis}$ is related to the Born cross section $\sigma_\text{born}$ according to the formula~\cite{kuraev}:
\begin{equation}
\label{eq:radcor}
	\sigma_\text{vis}(E) = \int\limits^1_0 F(x,E)\ \sigma_\text{born} \left( E \sqrt{1-x} \right) dx,
\end{equation}
where $F(x,E)$ is a function describing the probability of energy loss $xE/2$ due to radiation from the initial state.
This formula can be rewritten as
\begin{equation}
\label{eq:radcor1}
	\sigma_\text{vis}(E) = \sigma_\text{born}(E)(1+\delta_\text{rad}(E)),
\end{equation}
where $\delta_\text{rad}(E)$ is the radiative correction.

The visible cross section is obtained from the experimental data using to the formula:
\begin{equation}
\label{eq:viscs}
	\sigma_{\text{vis},i} = \frac{N_{{\rm exp},i}}{L_i \varepsilon_i},
\end{equation}
where $N_{{\rm exp},i}$ is the number of selected $\omega\pi^0$ events, $L_i$ is the integrated luminosity, and $\varepsilon_i$ is the detection efficiency for $i$-th energy point.
To obtain the experimental values of the Born cross section, the data on the visible cross section are fitted by the integral~(\ref{eq:radcor}), in which the theoretical model~(\ref{eq:borncs}) is used for the Born cross section, the parameters of which are determined from the fit.
Then, using the theoretical model, the radiative correction is calculated as
\begin{equation}
\label{eq:radcordef}
	\delta_\text{rad}(E) = \frac{ \sigma_\text{vis}(E) }{ \sigma_\text{born}(E) } - 1.
\end{equation}
The values of the Born cross section are calculated from $\sigma_{\text{vis},i}$ using Eq.~(\ref{eq:radcor1}).

The form factor in Eq.~(\ref{eq:borncs}) in the VMD model is parametrized as follows~\cite{snd2000a,kardopo13}:
\begin{equation}
\label{eq:formfactor}
	F_{\gamma\omega\pi}(E) = \frac{g_{\rho\omega\pi}}{f_\rho}
	\sum\limits_{i=0}^3 \frac{A_i M_i^2 e^{i \phi_i}}{M_i^2 - E^2 - iE\Gamma_i(E)},
\end{equation}
where the summation is over four isovector resonances $\rho(770)$, $\rho(1450)$, $\rho(1700)$, and $\rho(2150)$, and $M_i$, $\Gamma_i$, and $ \phi_i$ are masses, widths and phases of these resonances.
Coupling constants $f_\rho$ and $g_{\rho\omega\pi}$ are calculated from the decay widths \( \rho \to e^+e^- \) and \( \omega \to \pi^0\gamma \)~\cite{pdg}, respectively.
To describe the shape of the $\rho(770)$ resonance, the energy-dependent width is used
\begin{equation}
\label{eq:wrho}
	\Gamma_0(E) = \Gamma_0(M_0)\left( \frac{M_0}{E}\right)^{2}
		\left( \frac{E^2 - 4 m_\pi^2}{M_0^2 - 4 m_\pi^2} \right)^{3/2} +
		\frac{g_{\rho\omega\pi}^2}{4\pi} P_f(E),
\end{equation}
where the first term corresponds to the decay \( \rho \to \pi^+\pi^- \), and the second to the decay \( \rho \to \omega\pi^0 \).
For excited resonances, energy-independent widths are used.
The parameters of $\rho(770)$ are fixed at the Particle data group (PDG) values~\cite{pdg}, $M_0 = 775$~MeV and $\Gamma_0 =149.4$~MeV, and $\phi_0 \equiv 0$.
The parameters $A_0$, $A_1$, $A_2$, $A_3$, $\phi_1$, $\phi_2$, $\phi_3$, $M_1$, $M_3$, $\Gamma_1$ and $\Gamma_3$ are determined by the fit.
The data on the cross section \(e^+e^- \to \omega\pi^0 \) are weakly sensitive to the parameters of the $\rho(1700)$ resonance.
Its mass and width are varied in the fit near the PDG values with their Gaussian errors: \( M_2 = 1.72 \pm 0.02 \) GeV and \( \Gamma_2 = 0.25 \pm 0.10 \) GeV~\cite{pdg}.

To more accurately determine the contribution of the $\rho(770)$ resonance (parameter $A_0$), the SND data obtained at the VEPP-2M collider at energies below 1.02 GeV~\cite{snd2000b,snd2000a,snd2009} are added to the fit.
Some of these measurements were made in the \( \omega \to \pi^0\gamma \) channel.
For them, the cross section was recalculated using the ratio of the branching fractions \( B(\omega \to \pi^0 \gamma) / B(\omega \to \pi^+\pi^-\pi^0) = 0.0992 \pm 0.0023 \)~\cite{snd2016_pig} and the ratio of the phase spaces for the final states $\pi^+\pi^-\pi^0\pi^0$ and $\pi^0\pi^0\gamma$.

At the energy \( E > 1.9 \text{ GeV} \), the behavior of the Born cross section begins to be determined by the $\rho(2150)$ resonance observed in the BABAR and BESIII experiments~\cite{babar17,bes2021,qin2022}.
The parameters of this resonance cannot be entirely determined from our data.
To fix the $\rho(2150)$ parameters, the BABAR~\cite{babar17} data in the energy range above 1.5 GeV and the BESIII~\cite{bes2021} data below 2.5 GeV are added to the fit.
To take into account possible systematic shifts between measurements performed in different experiments, the BABAR and BESIII data are multiplied by the scale factors $S_\text{BABAR}$ and $S_\text{BESIII}$, which are determined from the fit.
The addition of the BABAR measurements is necessary because our and BESIII data do not overlap in energy.
It should be noted that the extraction of the $\omega\pi^0$ signal in the BESIII~\cite{bes2021} and BABAR~\cite{babar17} experiments is based on the fit to the $\pi^+\pi^-\pi^0$ mass spectrum with the sum of the $\omega$-resonance line shape and the nonresonant background distribution.
The $\omega$ line shape is described by the convolution of the Breit-Wigner distribution with the detector resolution function.
As can be seen from Fig.~\ref{fig:omlnshape} for such an approach, an adequate description of the energy dependence of the phase volume is the narrow resonance approximation.
Therefore, for inclusion to the fit and comparison with the SND data, the BESIII and BABAR measurements are multiplied by the ratio of the phase spaces shown in Fig.~\ref{fig:omlnshape} (right) as ``$\pi^+\pi^-\pi ^0$ ($m < 0.9$ GeV)'' and ``$p^3$''.
For BESIII data, this factor is approximately 1.02.
\begin{figure}
\centering
\includegraphics[width=\columnwidth]{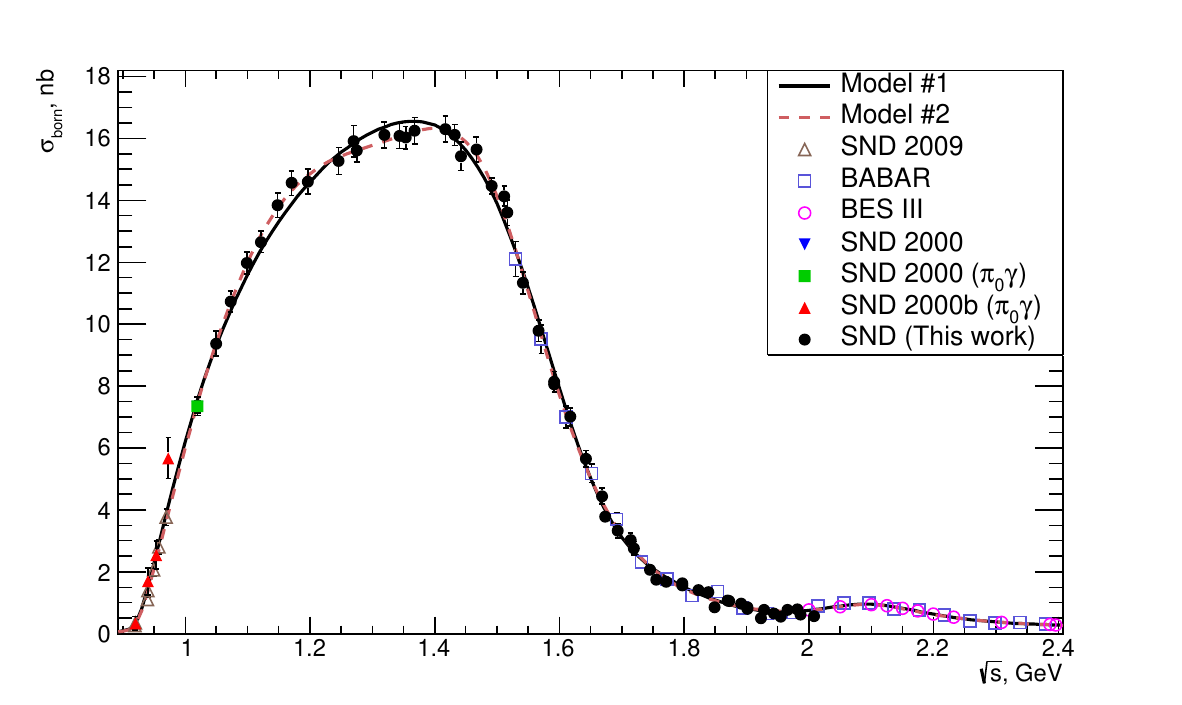}%
\caption{
	The Born cross section for the process \( e^+e^- \to \omega\pi^0 \to \pi^+\pi^-\pi^0\pi^0 \) in the energy range 0.9--2.4 GeV fitted with the VMD model.
	The data used in the fit from this work and other experiments: SND 2000~\cite{snd2000b,snd2000a}, SND 2009~\cite{snd2009}, BABAR~\cite{babar17}, and BESIII~\cite{bes2021}, are shown.
	The solid (dotted) curve represents the result of the fit with Model~1 (Model~2).
}
\label{fig:crossfit}
\end{figure}

\begin{table}
\caption{The parameters of the two models of the Born cross section obtained from the fit described in the text.}
\label{tab:rhofit}
\begin{ruledtabular}
\begin{tabular}{lcc}
\multicolumn{1}{c}{Param.} & \multicolumn{1}{c}{Model~1} & \multicolumn{1}{c}{Model~2} \\ \hline
$A_{0}$ & $0.894 \pm 0.006$ & $0.916 \pm 0.012$ \\
$A_{1}$ & $0.584 \pm 0.003$ & $0.164 \pm 0.003$ \\
$M_{1}$, GeV & $1.614 \pm 0.002$ & $1.523 \pm 0.004$ \\
$\Gamma_{1}$, GeV & $0.492 \pm 0.004$ & $0.368 \pm 0.006$ \\
$\phi_{1}$, rad & $3.106 \pm 0.004$ & $6.083 \pm 0.019$ \\
$A_{2}$ & $0.370 \pm 0.003$ & $0.002 \pm 0.001$ \\
$M_{2}$, GeV & $1.723 \pm 0.002$ & $\equiv 1.720$ \\
$\Gamma_{2}$, GeV & $0.371 \pm 0.003$ & $\equiv 0.250$ \\
$\phi_{2}$, rad & $0.300 \pm 0.006$ & $2.984 \pm 0.633$ \\
$A_{3}$ & $0.042 \pm 0.001$ & $0.005 \pm 0.000$ \\
$M_{3}$, GeV & $2.095 \pm 0.004$ & $2.088 \pm 0.006$ \\
$\Gamma_{3}$, GeV & $0.270 \pm 0.003$ & $0.211 \pm 0.008$ \\
$\phi_{3}$, rad & $5.820 \pm 0.020$ & $6.003 \pm 0.081$ \\
$A_{4}$ &  & $0.602 \pm 0.009$ \\
$M_{4}$, GeV &  & $1.183 \pm 0.006$ \\
$\Gamma_{4}$, GeV &  & $0.548 \pm 0.009$ \\
$\phi_{4}$, rad &  & $3.300 \pm 0.010$ \\
$S_\text{BABAR}$ & $1.028 \pm 0.022$ & $1.020 \pm 0.022$ \\
$S_\text{BESIII}$ & $0.884 \pm 0.025$ & $0.876 \pm 0.026$ \\
$\chi^2$ / ndf & $102.5\ /\ 90$ & $81.9\ /\ 86$ \\
\end{tabular}

\end{ruledtabular}
\end{table}

The parameters obtained from the fit in the model described above are listed in the ``Model~1'' column of Table~\ref{tab:rhofit}.
The result of this fit is shown in Fig.~\ref{fig:crossfit} as a solid curve.
Although the fit has an acceptable $\chi^2 / \text{ndf} = 102 / 90$, where ndf is the number of degrees of freedom, the description of the data in the 1.1--1.3 GeV region does not seem entirely satisfactory.
For example, 5 points in a row in the range 1.07--1.17 GeV lie above the fitting curve.
This may be due to incorrect description of the tail of the subthreshold resonance $\rho(770)$ and the shape of the $\rho(1450)$ resonance in Model~1.
Within the approach used, the problem is solved by adding one more resonance with a mass of about 1.1 GeV (Model~2).
In this model, the mass and width of $\rho(1700)$ have been exactly fixed at their PDG values.
The fit results are shown in Table~\ref{tab:rhofit} in the column ``Model~2''.
The fitting curve is shown in Fig.~\ref{fig:crossfit} as a dotted line.
Model~2 has a substantially better $\chi^2 / \text{ndf} = 82 / 86$.

The difference of the coefficients $S_\text{BABAR}$ and $S_\text{BESIII}$ from unity and from each other is within the systematic uncertainties near 2 GeV, which are 10\% for BABAR~\cite{babar17}, 8--12\% for BESIII~\cite{bes2021} and 6.6\% for SND.
Here, the systematic uncertainty associated with interference, which is common for all three measurements, is excluded from the SND error.

Model~1 is used to obtain the Born cross section and radiative correction.
The difference in the radiative correction between Models~1 and 2 is taken as an estimate of its systematic error.
The obtained cross section values are listed in Table~\ref{tab:crossfit} and shown in Fig.~\ref{fig:crossfit}.

The procedure for determining the Born cross section described above is actually a numerical solution of the integral equation (\ref{eq:radcor}), in which a theoretical model is used for regularization.
The obtained values of the Born cross section are correlated.
To determine the covariance matrix, a series of 10 thousand pseudoexperiments is used.
Data is generated according to a theoretical model with variances corresponding to the statistical errors in the number of signal events.
The statistical uncertainties of the Born cross section in Table~\ref{tab:crossfit} and in Fig.~\ref{fig:crossfit} correspond to the diagonal elements of the covariance matrix.
The full covariance matrix is given in the supplementary materials.

\begin{table*}
\caption{Contributions to the systematic uncertainty in the Born cross section from different sources.}
\label{tab:syst}
\begin{ruledtabular}
\begin{tabular}{lrrrrr}
$E$, MeV & 1000 -- 1200 & 1200 -- 1400 & 1400 -- 1600 & 1600 -- 1800 & 1800 -- 2000 \\\hline
Luminosity & 2.0 \% & 2.0 \% & 2.0 \% & 2.0 \% & 2.0 \% \\
Eff. correction & 2.1 \% & 2.1 \% & 2.1 \% & 2.1 -- 2.3 \% & 2.3 \% \\
Background & 0.0 -- 0.2 \% & 0.0 -- 0.1 \% & 0.0 -- 0.2 \% & 0.2 -- 0.6 \% & 0.3 -- 0.6 \% \\
Interference & 0.5 -- 1.1 \% & 1.1 -- 2.4 \% & 2.4 -- 4.1 \% & 4.1 -- 9.4 \% & 8.3 -- 13.3 \% \\
Model & 0.4 \% & 0.4 \% & 0.4 -- 0.9 \% & 0.9 -- 2.4 \% & 2.4 -- 3.9 \% \\
Rad. correction & 0.0 -- 0.3 \% & 0.0 -- 0.4 \% & 0.0 -- 0.7 \% & 0.4 -- 3.4 \% & 0.1 -- 4.3 \% \\\hline
Total & 3.0 -- 3.2 \% & 3.2 -- 3.8 \% & 3.8 -- 5.1 \% & 5.1 -- 10.7 \% & 9.7 -- 14.2 \% \\
\end{tabular}

\end{ruledtabular}
\end{table*}

\begin{figure*}
\centering
\includegraphics[width=0.5\textwidth]{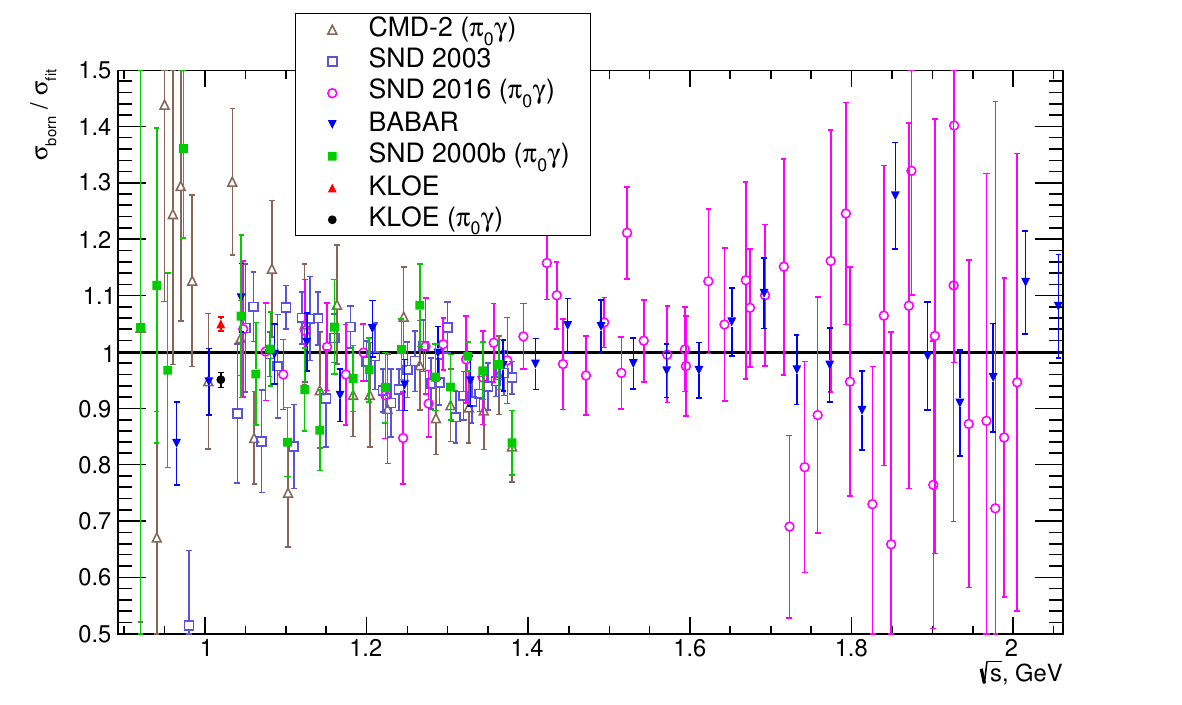}
\caption{
	The ratio of the \( e^+e^- \to \omega\pi^0 \to \pi^+\pi^-\pi^0\pi^0 \) Born cross sections measured in the CMD-2~\cite{cmd03}, SND 2003~\cite{snd2003}, SND 2016~\cite{snd2016}, KLOE~\cite{kloe08} and BABAR~\cite{babar17} experiments to the fit result in Model~1.
}
\label{fig:crossfitnorm}
\end{figure*}


The following effects, discussed in previous sections, contribute to the systematic uncertainty in the cross section.
The accuracy of the luminosity measurement is 2\%.
The uncertainty of the efficiency corrections is 2.1--2.3\% depending on the energy.
The model uncertainty arising from background subtraction from non-$\omega\pi^0$ mechanisms of the process \( e^+e^- \to \pi^+\pi^-\pi^0\pi^0 \) varies in the range 0.4--3.9\%.
The systematic uncertainty coming from the subtraction of background processes other than \( e^+e^- \to \pi^+\pi^-\pi^0\pi^0 \) is less than 0.6\%.
The uncertainty coming from the interference of $\omega\pi^0$ with other intermediate mechanisms of the \( e^+e^- \to \pi^+\pi^-\pi^0\pi^0 \) process varies from 0.5\% at 1 GeV to 13.3\% at 2 GeV.
The model uncertainty of the radiative correction does not exceed 4.3\%.
The systematic uncertainties from different sources in five energy ranges and their quadratic sum are listed in Table~\ref{tab:syst}.

\begin{table*}[p]
\caption{
	The c.~m. energy ($E$), selected number of $\omega\pi^0$ events ($N_{\rm exp}$), integrated luminosity ($L$), detection efficiency ($\varepsilon$), Born cross section for the process \( e^+e^- \to \omega\pi^0 \to \pi^+\pi^-\pi^0\pi^0 \) ($\sigma_\text{born}$) with constraint $m_{3\pi}^\text{truth} < 0.9$ GeV, and radiative correction ($\delta_\text{rad}$).
	For the number of events, the statistical error is quoted.
	For the Born cross section, the first error is statistical, and second is systematic.
}
\label{tab:crossfit}
\begin{ruledtabular}
\def\arraystretch{0.6}
\begin{tabular}{r r r r r r r}
  $E$, MeV & \multicolumn{1}{c}{$N_\text{exp}$} & \multicolumn{1}{c}{$IL\text{, nb}^{-1}$} & \multicolumn{1}{c}{$\varepsilon$} & \multicolumn{1}{c}{$\sigma_\text{born}$, nb} & $\delta_\text{rad}$ \\ \hline
  1049.7 & $1048.2 \pm  45.4$ &   420 & 0.296 & $  9.37 \pm 0.42 \pm 0.28$ & $ -0.100$ \\
  1072.7 & $1618.2 \pm  49.2$ &   554 & 0.299 & $ 10.74 \pm 0.34 \pm 0.33$ & $ -0.092$ \\
  1099.0 & $1780.1 \pm  51.3$ &   560 & 0.290 & $ 11.98 \pm 0.37 \pm 0.37$ & $ -0.083$ \\
  1121.4 & $2078.4 \pm  54.7$ &   543 & 0.328 & $ 12.65 \pm 0.35 \pm 0.39$ & $ -0.077$ \\
  1148.4 & $2032.9 \pm  56.3$ &   495 & 0.319 & $ 13.84 \pm 0.40 \pm 0.43$ & $ -0.071$ \\
  1170.6 & $2429.1 \pm  63.1$ &   547 & 0.327 & $ 14.56 \pm 0.39 \pm 0.46$ & $ -0.066$ \\
  1196.7 & $2440.0 \pm  64.6$ &   570 & 0.312 & $ 14.61 \pm 0.40 \pm 0.46$ & $ -0.061$ \\
  1246.1 & $2245.1 \pm  62.9$ &   481 & 0.322 & $ 15.28 \pm 0.45 \pm 0.51$ & $ -0.053$ \\
  1270.4 & $2503.4 \pm  76.9$ &   511 & 0.324 & $ 15.91 \pm 0.50 \pm 0.55$ & $ -0.050$ \\
  1275.3 & $3511.7 \pm  79.1$ &   744 & 0.318 & $ 15.61 \pm 0.37 \pm 0.54$ & $ -0.049$ \\
  1319.1 & $2782.8 \pm  69.8$ &   569 & 0.317 & $ 16.12 \pm 0.42 \pm 0.59$ & $ -0.043$ \\
  1343.5 & $2966.9 \pm  71.6$ &   599 & 0.321 & $ 16.08 \pm 0.41 \pm 0.59$ & $ -0.040$ \\
  1353.3 & $4129.6 \pm  86.3$ &   845 & 0.317 & $ 16.03 \pm 0.36 \pm 0.59$ & $ -0.038$ \\
  1368.1 & $3076.9 \pm  75.0$ &   625 & 0.314 & $ 16.25 \pm 0.41 \pm 0.61$ & $ -0.036$ \\
  1417.3 & $3011.0 \pm  74.6$ &   601 & 0.316 & $ 16.30 \pm 0.43 \pm 0.62$ & $ -0.026$ \\
  1432.1 & $5100.3 \pm  97.7$ &  1034 & 0.313 & $ 16.11 \pm 0.34 \pm 0.62$ & $ -0.023$ \\
  1442.0 & $2242.6 \pm  63.9$ &   474 & 0.313 & $ 15.42 \pm 0.46 \pm 0.60$ & $ -0.020$ \\
  1466.8 & $3027.1 \pm  74.9$ &   623 & 0.314 & $ 15.65 \pm 0.41 \pm 0.62$ & $ -0.012$ \\
  1491.7 & $6817.8 \pm 111.6$ &  1520 & 0.310 & $ 14.47 \pm 0.26 \pm 0.60$ & $ -0.001$ \\
  1511.7 & $3934.1 \pm  87.0$ &   925 & 0.298 & $ 14.14 \pm 0.34 \pm 0.61$ & $  0.010$ \\
  1516.7 & $2121.4 \pm  62.5$ &   511 & 0.301 & $ 13.60 \pm 0.41 \pm 0.59$ & $  0.013$ \\
  1541.7 & $2006.0 \pm  62.8$ &   575 & 0.298 & $ 11.33 \pm 0.37 \pm 0.52$ & $  0.032$ \\
  1566.8 & $1613.2 \pm  57.0$ &   540 & 0.289 & $  9.79 \pm 0.36 \pm 0.48$ & $  0.057$ \\
  1592.0 & $2380.1 \pm  70.5$ &  1021 & 0.265 & $  8.05 \pm 0.26 \pm 0.41$ & $  0.091$ \\
  1592.0 & $1130.0 \pm  46.4$ &   464 & 0.275 & $  8.13 \pm 0.35 \pm 0.42$ & $  0.091$ \\
  1617.3 & $1156.2 \pm  48.1$ &   548 & 0.265 & $  7.01 \pm 0.31 \pm 0.39$ & $  0.136$ \\
  1642.7 & $ 843.0 \pm  40.0$ &   506 & 0.247 & $  5.65 \pm 0.28 \pm 0.39$ & $  0.192$ \\
  1668.2 & $ 636.7 \pm  37.5$ &   485 & 0.235 & $  4.44 \pm 0.28 \pm 0.32$ & $  0.259$ \\
  1673.3 & $ 999.4 \pm  49.3$ &   909 & 0.228 & $  3.79 \pm 0.20 \pm 0.28$ & $  0.273$ \\
  1693.7 & $ 467.3 \pm  32.7$ &   489 & 0.216 & $  3.33 \pm 0.25 \pm 0.27$ & $  0.329$ \\
  1714.2 & $ 419.2 \pm  32.0$ &   501 & 0.200 & $  3.02 \pm 0.24 \pm 0.26$ & $  0.384$ \\
  1719.4 & $ 418.8 \pm  32.5$ &   542 & 0.201 & $  2.75 \pm 0.23 \pm 0.24$ & $  0.397$ \\
  1745.1 & $ 294.7 \pm  26.8$ &   527 & 0.186 & $  2.07 \pm 0.20 \pm 0.19$ & $  0.457$ \\
  1755.5 & $ 422.2 \pm  34.6$ &   941 & 0.174 & $  1.75 \pm 0.15 \pm 0.17$ & $  0.479$ \\
  1771.0 & $ 218.9 \pm  23.5$ &   503 & 0.171 & $  1.68 \pm 0.19 \pm 0.17$ & $  0.512$ \\
  1796.9 & $ 150.4 \pm  18.2$ &   410 & 0.149 & $  1.56 \pm 0.19 \pm 0.17$ & $  0.571$ \\
  1796.9 & $ 379.1 \pm  30.0$ &  1018 & 0.145 & $  1.64 \pm 0.14 \pm 0.18$ & $  0.571$ \\
  1823.0 & $ 169.5 \pm  18.4$ &   530 & 0.139 & $  1.40 \pm 0.16 \pm 0.15$ & $  0.640$ \\
  1838.6 & $ 262.0 \pm  26.3$ &   910 & 0.127 & $  1.34 \pm 0.14 \pm 0.14$ & $  0.687$ \\
  1849.1 & $  81.5 \pm  16.9$ &   436 & 0.127 & $  0.86 \pm 0.18 \pm 0.09$ & $  0.717$ \\
  1870.1 & $ 147.0 \pm  20.9$ &   672 & 0.115 & $  1.06 \pm 0.16 \pm 0.10$ & $  0.782$ \\
  1872.2 & $ 199.3 \pm  20.8$ &   939 & 0.113 & $  1.05 \pm 0.12 \pm 0.10$ & $  0.788$ \\
  1891.2 & $ 121.5 \pm  16.4$ &   620 & 0.109 & $  0.97 \pm 0.14 \pm 0.10$ & $  0.841$ \\
  1901.7 & $  81.7 \pm  13.0$ &   498 & 0.103 & $  0.85 \pm 0.14 \pm 0.08$ & $  0.864$ \\
  1901.7 & $ 148.1 \pm  17.5$ &   959 & 0.103 & $  0.81 \pm 0.10 \pm 0.08$ & $  0.864$ \\
  1922.9 & $  62.2 \pm  13.4$ &   677 & 0.098 & $  0.50 \pm 0.11 \pm 0.05$ & $  0.889$ \\
  1928.2 & $  87.8 \pm  12.7$ &   624 & 0.098 & $  0.76 \pm 0.11 \pm 0.08$ & $  0.890$ \\
  1944.1 & $ 105.2 \pm  17.6$ &   935 & 0.092 & $  0.65 \pm 0.11 \pm 0.07$ & $  0.878$ \\
  1954.8 & $  38.0 \pm   8.6$ &   432 & 0.086 & $  0.55 \pm 0.13 \pm 0.06$ & $  0.858$ \\
  1965.4 & $  90.0 \pm  14.0$ &   745 & 0.086 & $  0.77 \pm 0.12 \pm 0.08$ & $  0.828$ \\
  1981.5 & $  61.9 \pm  12.6$ &   515 & 0.086 & $  0.79 \pm 0.16 \pm 0.09$ & $  0.768$ \\
  1986.8 & $  60.8 \pm  12.0$ &   644 & 0.088 & $  0.62 \pm 0.13 \pm 0.07$ & $  0.746$ \\
  2008.3 & $  46.2 \pm   9.6$ &   587 & 0.084 & $  0.57 \pm 0.12 \pm 0.08$ & $  0.650$ \\
\end{tabular}

\end{ruledtabular}
\end{table*}

\section{Conclusion}

The cross section of the process \( e^+e^- \to \omega\pi^0 \to \pi^+\pi^-\pi^0\pi^0 \) has been measured in the SND experiment at the VEPP-2000 $e^+e^-$ collider in the energy range 1.05--2.00 GeV.
The statistical uncertainty of the measurement is 2--23\%, while the systematic uncertainty is 3.0--14.2\%.

Comparison of the current SND measurement with results of previous experiments is demonstrated in Fig.~\ref{fig:crossfitnorm}, where the ratio of the CMD-2~\cite{cmd03}, SND 2003~\cite{snd2003}, SND 2016~\cite{snd2016}, KLOE~\cite{kloe08} and BABAR~\cite{babar17} data to the result of the fit to our data in Model~1 is shown.
All the measurements are in good agreement with each other in the region under study 1.05--2.00 GeV.
The most accurate previous measurements were carried out in the BABAR experiment in the $\pi^+\pi^-\pi^0\pi^0$ channel~\cite{babar17}, and in the SND experiment in the $\pi^0\pi^0\gamma$ channel~\cite{snd2016}.
The BABAR data have a statistical accuracy comparable to our measurement up to 1.5 GeV and slightly better at higher energies, but worse (about 10\%) systematic uncertainty.
Moreover, BABAR~\cite{babar17} did not study the model uncertainty associated with the interference between different states that contribute to the process \( e^+e^- \to \pi^+\pi^-\pi^0\pi^0 \).
Our estimate of this uncertainty given in Table~\ref{tab:syst} can be applied to the BABAR data as well.
In the SND measurement in the $\pi^0\pi^0\gamma$ channel, on the contrary, the systematic uncertainty is smaller (2.7--5.2\%), while the statistical uncertainty is larger.
In general, our new measurement can be characterized as the most accurate in the energy region under study at the present time.

The maximum contribution to the systematic uncertainty of our measurement above 1.2 GeV comes from the model error associated with the interference of the $\omega\pi^0$ mechanism with other mechanisms of the process \( e^+e^- \to \pi^+\pi^-\pi^0\pi^0 \), as well as from the uncertainty related to background subtraction from these mechanisms.
Reduction of these uncertainties is possible only with the use of amplitude analysis of the process \( e^+e^- \to \pi^+\pi^-\pi^0\pi^0 \) preferably together with the process \(e^+e^- \to \pi^+\pi^-\pi^+\pi^- \), as it was done in Ref.~\cite{kozyrev}.
However, more promising, in our opinion, is the increase in statistics in the SND experiment for the \( e^+e^- \to \omega\pi^0 \to \pi^0\pi^0\gamma \) channel.
In this channel, the model uncertainty associated with interference with the intermediate state $\rho^0\pi^0$, can be significantly reduced taking into account the results of the Dalitz plot analysis in the process \( e^+e^- \to \pi^+\pi^-\pi^0 \)~\cite{snd_3pi_3}.

The measured cross section of the process \( e^+e^- \to \omega\pi^0 \to \pi^+\pi^-\pi^0\pi^0 \) together with SND data~\cite{snd2000b,snd2000a,snd2009} below 1.05~GeV and the BABAR~\cite{babar17} and BESIII~\cite{bes2021} data above 1.5~GeV is well described by the vector dominance model with four known resonances of the $\rho$ family.
The obtained parameters of the resonances are given in Table~\ref{tab:rhofit} (Model~1).

\section{Acknowledgments}

We thank \fbox{S.~I. Eidelman} and A.~I. Milstein for the fruitful discussions.
This work is supported by the Ministry of Education and Science of the Russian Federation.


\end{document}